\documentclass[fleqn,usenatbib]{mnras}

\usepackage{newtxtext,newtxmath}

\usepackage[T1]{fontenc}
\usepackage{ae,aecompl}

\usepackage[toc,page]{appendix}
\usepackage{float}
\usepackage{stfloats}

\usepackage{graphicx}	
\usepackage{amsmath}	
\usepackage{caption}

\title[The SAMI -- Fornax Dwarfs Survey II]{The SAMI -- Fornax Dwarfs Survey II: The Stellar Mass Fundamental Plane and the Dark Matter fraction of Dwarf Galaxies}

\author[F.S. Eftekhari et al.]{
F. Sara Eftekhari$^{1}$,
Reynier F. Peletier$^{1}$\thanks{E-mail: peletier@astro.rug.nl},
Nicholas Scott$^{2,3}$,
Steffen Mieske$^{4}$,
\newauthor
Joss Bland-Hawthorn$^{2,3}$,
Julia J. Bryant$^{2,3}$,
Michele Cantiello$^5$,
Scott M. Croom$^{2,3}$,
\newauthor
Michael J. Drinkwater$^{6}$,
J\'esus Falc\'on-Barroso$^{7,8}$,
Michael Hilker$^{9}$,
Enrichetta Iodice$^{10}$,
\newauthor
Nicola R. Napolitano$^{10}$, 
Marilena Spavone$^{10}$,
Edwin A. Valentijn$^{1}$,
Glenn van de Ven$^{11}$
\newauthor
and Aku Venhola$^{12,1}$
\newauthor
\\
$^{1}$Kapteyn Institute, University of Groningen, Landleven 12, 9747, AD, Groningen, The Netherlands\\
$^{2}$Sydney Institute for Astronomy, School of Physics, A28, The University of Sydney, NSW, 2006, Australia\\
$^{3}$ARC Centre of Excellence for All Sky Astrophysics in 3 Dimensions (ASTRO 3D)\\
$^{4}$European Southern Observatory, Alonso de Cordova 3107, 7630355 Vitacura, Santiago, Chile\\
$^5$INAF Osservatorio Astronomico di Teramo, via Maggini, I-64100, Teramo, Italy\\
$^{6}$School of Mathematics and Physics, The University of Queensland, Brisbane, QLD 4072, Australia\\
$^{7}$Instituto de Astrof\'isica de Canarias, Calle V\'ia L\'actea s/n, E-38205 La Laguna, Tenerife, Spain\\ 
$^{8}$Departamento de Astrof\'isica, Universidad de La Laguna (ULL), E-38206 La Laguna, Tenerife, Spain\\
$^{9}$ESO, European Southern Observatory, Karl-Schwarzschild-Str 2, D-85748 Garching bei M{\"u}nchen, Germany\\
$^{10}$INAF - Astronomical Observatory of Capodimonte, via Moiariello 16, I-80131 Napoli, Italy\\
$^{11}$Department of Astrophysics, University Vienna, T{\"u}rkenschanzstrasse 17, A-1180 Wien, Austria\\
$^{12}$Space physics and astronomy research unit, University of Oulu, Pentti Kaiteran katu 1, FI-90014 Oulu, Finland\\
}
\date{Accepted XXX. Received YYY; in original form ZZZ}

\pubyear{2021}

\begin{document}
\label{firstpage}
\pagerange{\pageref{firstpage}--\pageref{lastpage}}
\maketitle

\begin{abstract}
We explore the kinematic scaling relations of 38 dwarf galaxies in the Fornax Cluster using observations from the SAMI integral field spectrograph. We focus on the Fundamental Plane (FP), defined by the physical properties of the objects (scale length, surface brightness and velocity dispersion) and the Stellar Mass (Fundamental) Plane, where surface brightness is replaced by stellar mass, and investigate their dynamical-to-stellar-mass ratio. 
We confirm earlier results that the Fornax dEs are significantly offset above the FP defined by massive, hot stellar systems. For the Stellar Mass (Fundamental) Plane, which shows much lower scatter, we find that young and old dwarf galaxies lie at about the same distance from the plane, all with comparable scatter. We introduce the perpendicular deviation of dwarf galaxies from the Stellar Mass Plane defined by giant early-types as a robust estimate of their DM fraction, and find that the faintest dwarfs are systematically offset above the plane, implying that they have a higher dark matter fraction. This result is confirmed when estimating the dynamical mass of our dEs using a virial mass estimator, tracing the onset of dark matter domination in low mass stellar systems. 
We find that the position of our galaxies on the Stellar Mass FP agrees with the galaxies in the Local Group. This seems to imply that the processes determining the position of dwarf galaxies on the FP depend on the environment in the same way, whether the galaxy is situated in the Local Group or in the Fornax Cluster.

 
\end{abstract}




\section{Introduction}

It is known that the properties and evolutionary history of a galaxy mainly depend on its local environment and total mass \citep{Disney:2008}. A very effective way to study the properties and evolutionary history is to use scaling relations between the internal properties of the galaxies, such as the FP \citep{Djorgovski:1985, Dressler:1987, Toloba:2012, Kormendy:2009}, the Faber-Jackson relation \citep{Faber:1976}, and the color-magnitude relation \citep{Sandage:1978}. Stellar and dynamical masses, or closely associated quantities, are essential factors in determining when and how the star formation in galaxies stopped. Based on \citet{Peng:2010}, star formation halts in two ways. In the first scenario the halo of a galaxy with M$_{*} >> 10^{9.5}$ M$_\odot$ prevents formation of the dense gas required for star formation. In the second, star formation is stopped by the environment, a factor outside of the halo, which impacts galaxies of all masses \citep{Silk:2012}. In this paper, where we discuss the properties of dwarf galaxies, environmental quenching is the main process, for low-mass objects in a dense environment.

\subsection{Current ideas of the formation of dwarf galaxies}

\citet{Dressler:1980} found a strong relation between the morphology of a galaxy and its environment, showing that in the local Universe, red early-type galaxies mostly appear in dense regions, while blue late-type spiral galaxies are mainly found in the low-density field. Although a strong relation for massive galaxies, this morphology-density relation manifests itself even more strongly for low mass galaxies, such as dwarf elliptical galaxies, which are the dominant population in number in large groups and clusters \citep{Ferguson:1989, Binggeli:1990, Binggeli:1988, Moore:1998, Boselli:2008, Serra:2012}. The strong morphology-density relation in nearby galaxy clusters \citep{Cappellari:2016} indicates that for galaxies inside clusters environmental effects, such as ram-pressure stripping \citep{Lin:1983}, harassment \citep{Moore:1998}, or starvation \citep{Larson:1980} are very important. Low mass objects, with their weak potential wells, are the most vulnerable to these strong environmental effects. 

In this paper we provide data, which can be used to understand which one of these processes is most important in the evolution of cluster dwarf galaxies. In the case of ram-pressure stripping, in clusters with hot gas, especially near their centers, the cold gas component of galaxies is partly removed during each crossing by the hot X-ray emitting gas of the cluster \citep{Gunn:1972, Boselli:2014}. When ram-pressure stripping is intense, it can remove the atomic gas disk \citep{Boselli:2008}, however, when it is not strong enough, it still strips away the warm ionized halo gas of the galaxy \citep{Font:2008, McCarthy:2008, Bekki:2009}. This process is known as starvation/strangulation, which halts gas accretion, stops the cooling process of the hot halo, and forces the galaxy to use up its last disk gas \citep{Larson:1980}. Removal of gas by ram pressure stripping and starvation causes slow changes in the morphology of the stellar component by making the galaxy structure smoother. Also, the cold gas that is falling into the central regions of the galaxy results in central star formation and as a result dwarf galaxies with blue centers \citep{Lisker:2006, Hamraz:2019} are formed. Ram-pressure stripping can lead to spectacular HI tails that have been blown out of the in-falling late-type galaxies \citep{Kenney:2004, Jaffe:2018}. As a result of ram-pressure effects, field galaxies contain more gas than cluster galaxies \citep{Solanes:2001, Serra:2012}. If the star formation change is significant, it can create a dichotomy in color-magnitude diagrams, in which the majority of gas-poor early-type galaxies reside along the red sequence, and a blue cloud, consisting of gas-rich late-type galaxies, lies below the line. Ram pressure stripping and strangulation do not disturb the stellar kinematics of a galaxy significantly.

Processes that do affect the stellar kinematics are harassment, mergers, and large tidal interactions, which all heat up stellar population of the galaxy. 
While mergers, violent types of galaxy interactions, are relatively common for massive galaxies, mergers involving dwarf galaxies are mostly minor, since otherwise the dwarf galaxies, with their much lower stellar densities, would be destroyed.
In clusters with large masses ($\sim10^{14-15}M_{\odot}$) repeated tidal interactions occur between  galaxies at high relative velocities, and between galaxies and the cluster potential. This effect, which is called harassment, influences the stellar mass, the dark matter, and the gas content of the galaxy and can alter the morphology of objects \citep{Mastropietro:2005, Fujita:1998}. As their outskirts are ripped off, galaxies become more compact, and their size will change, followed by their kinematics, with the strength of mass loss from harassment being entirely dependent on the orbital velocity of the galaxy \citep{Moore:1998, Mastropietro:2005, Smith:2015, Smith:2013, Smith:2010}. 

\subsection{Dwarf Galaxies in the Fornax Cluster}

In this paper we discuss scaling relations in the Fornax Cluster, with a distance of 19.7 Mpc (distance modulus of 31.51 \citep{Blakeslee:2009} the second closest galaxy cluster. Compared to the Virgo cluster, the Fornax cluster is dynamically more evolved in its central regions, but less well studied. It is smaller and denser, with a larger fraction of early-type galaxies in its central regions (see e.g., \citet{Iodice:2019b}. The Fornax cluster, with the central mean recession velocity of $1442\pm20$ $km^{-1}$ \citep{Maddox:2019}, a virial mass of $M=7\times10^{13}M_{\odot}$ and a virial radius of $\sim0.7$ Mpc \citep{Drinkwater:2001} is located in the southern sky in the Fornax-filament of the cosmic web \citep{Nasonova:2011} and contains hundreds of faint galaxies $M_B>-18$ \citep{Ferguson:1989, Venhola:2018}. While both internal and environmental mechanisms contribute to the cessation of star formation in the massive galaxies, environmental processes dominate for low-mass galaxies. 

The availability of our new optical catalog of about 600 dwarf galaxies in the Fornax Cluster, the FDSDC (Fornax Deep Survey Dwarf Catalogue), based on the Fornax Deep Survey (Venhola et al. 2018, Iodice et al. 2016), in u,g,r and i, makes it possible to study the effects of various physical processes on dwarf galaxies in clusters. This survey was published more or less at the same time as the NGFS (Next Generation Fornax Survey) (Eigenthaler et al. 2018), which presented a similar quality sample in a smaller area. For 33 bright galaxies in the Fornax Cluster MUSE IFU data have been obtained and studied as part of the F3D survey \citep{Sarzi:2018}. Also, recently, surveys are becoming available studying the ISM in the cluster. A pointed survey of the CO molecular gas in dwarf galaxies, based on the presence of dust from Herschel maps was published by Zabel et al. (2019). Recently, the first HI survey of the cluster was published by Loni et al. (2021), using ASKAP. More sensitive and higher resolution data are planned with MeerKAT by \citet{Serra:2016}. In the X-ray, there is the survey of \citet{Paolillo:2002}, which only covers the very center of the cluster. The dynamically evolved center of the cluster and its evolved galaxy population is a signature that galaxies have spent a few Gyrs in the cluster environment and traveled a few times through the cluster center \citep{Drinkwater:2001} and consequentially significant intracluster stars exist \citep{Pota:2018, Iodice:2017, Chaturvedi:2022, Napolitano:2022}. 

\subsection{The SAMI-Fornax Dwarf Survey}

Using integral field spectroscopy, we aim to understand the structure and formation of dwarf galaxies compared to their giant counterparts and the influence of the environment on the formation of dwarf galaxies. \citet{Binggeli:1984} defined dwarf galaxies to be  galaxies with ${\rm M_B} > -18$. Roughly consistent with their definition, our dwarf sample has absolute magnitudes in the $r$-band fainter than -19 mag. In this study, we do not include compact ellipticals, such as M32, or Ultra Compact Dwarfs (UCDs), but discuss all other objects discussed in \citet{Kormendy:2012} and \citet{Kormendy:2009}.
Here we present paper II of the SAMI-Fornax Dwarf Survey (referred to here as DSAMI), in which we present high-resolution spectra (R$\sim$5000) of the largest sample of low-mass ($10^{7}-10^{8.5}$) galaxies in a cluster to date. In the first paper \citep{Scott:2020}, the spatially resolved stellar radial velocity and velocity dispersion maps, together with their specific stellar angular momentum, $\lambda_R$, were presented. The relation between the specific angular momentum and stellar mass, as well as cluster-centric radius is studied in paper I.  We find that $\lambda_R$ decreases with mass from a maximum of $\sim$ 0.7-0.8 at around $M_* \approx 10^{10} M_\odot$ to much lower values of $\sim$ 0.2-0.4 for masses around $M_* \approx 10^{8} M_\odot$, implying that early-type dwarf galaxies are not supported by rotation. This results is comparable with the fact that the kinematics of the CALIFA survey show  that the latest-type spirals, possible progenitors of cluster dwarfs, are barely rotationally supported \citep{Falcon-Barroso:2019}. These are of lower mass than the other objects in CALIFA, but still about a factor 10 more massive than the brightest dwarfs in this paper.  Also, a modest dependence of $\lambda_R$ of dE galaxies on cluster-centric radius is reported, similar to findings of \citet{Toloba:2015}. These results are consistent with the formation scenario where dEs form from low mass spiral galaxies that have their gas stripped by cluster processes  such as ram-pressure stripping, consistent with the findings of \citet{Koleva:2014}. 

\subsection{Scaling Relations for Dwarf Galaxies}

In this second paper, we investigate the position of dwarf galaxies on scaling relations, such as the FP, and the relation between luminous and dark matter. We introduce a special kind of FP, the Stellar Mass Fundamental Plane, which allows us to correct for the effect of stellar populations. By replacing surface brightness by stellar luminosity density, or in practice stellar mass, the scaling relation is insensitive to stellar population differences, and can be used to investigate other properties, such as dark matter.

The behavior of dwarf galaxies on these scaling relations has been investigated in several papers such as \citet{Graham:2003, Geha:2003, deRijcke:2005, Kormendy:2009, Toloba:2012}. Kinematic scaling relations can be used to study the connection between properties of dwarfs and massive early-types and the dark matter content in them. Based on scaling relations, some works \citep{Kormendy:2009} consider dwarf and giant early-types as two separate populations while studies such as \citet{Graham:2003} find continuity in their physical properties.
\citet{Cappellari:2006}, through a virial mass estimate, find that baryons dominate the mass within the half-light radii of giant elliptical galaxies. In contrast, \citet{Wolf:2010} show that the situation for dwarf spheroidals and ultra-faint galaxies is very different, with large M$_{dyn}$ /L values within their half-light radii. $\Lambda$CDM based galaxy formation models (see the review of \citet{Bullock:2017}) reproduce this trend. However, the high dark matter values of \citet{Wolf:2010} are based only on Local Group galaxies. It is not known whether objects outside the Local Group follow the same trend.
\citet{deRijcke:2005} and \citet{Toloba:2012} claim that dwarf ellipticals are offset with respect to giant ellipticals on the FP, indicating a slightly larger mass-to-light ratio. \citet{Penny:2015} find that the mass-to-light ratio of dwarfs in groups is somewhat higher than that in the Virgo cluster.\\
In this paper, we study the FP for a large sample of dwarfs, extending to considerably fainter galaxies than has been done before in the cluster environment, to determine dark matter fractions for objects overlapping in mass with galaxies in the Local Group. For example, we can test whether the the structure of cluster galaxies is similar to those in the Local Group, i.e., whether the cluster environment creates dwarf galaxies that are different from those nearby.
The Fundamental Plane concept had not changed much since its discovery more than 30 years ago. By using stellar mass instead of absolute magnitude (see \citet{Hyde:2009}), we introduce a new plane, in which contamination by stellar population effects, such as deviations caused by the presence of young stellar populations, are minimized, causing a smaller scatter that is easier to model and understand. 

The paper is organized as follows: In section 2, we describe the SAMI-Fornax IFU data set and compare it with previous observations. In section 3, we explain the way we determine the galaxy parameters from their 3D cubes. Section 4 contains the scaling relations, the FP and the Stellar Mass Fundamental Plane. We also discuss the dark matter of dwarf galaxies. Section 5 discusses these results and their connection to formation scenarios of dwarf galaxies, followed by a conclusion section.

\begin{table*}
\centering

\caption{SAMI -- Fornax Dwarf Survey sample. For a full description of the sample - see Paper 1. All tabulated values for the dwarfs are from the FDS catalogue \citep{Venhola:2018,Venhola:2019}. In column (7) magnitudes in $r$ are given inside an effective radius. The surface brightness in column (8) is calculated using equation (2). }
\begin{tabular}{|rccccrclrclrclcccc|}
\hline
FCC & FDS & year & RA & Dec & \multicolumn{3}{c}{R$_e$} & \multicolumn{3}{c}{$M_{r,e}$} & \multicolumn{3}{c}{$\mu_{r,e}$}  & Axis & S\'ersic  & $g-r$ &$u-g$ \\ 
ID & ID &  & (deg) & (deg) & \multicolumn{3}{c}{(arcsec)} & \multicolumn{3}{c}{(mag)} & \multicolumn{3}{c}{($mag/{arcsec^2}$)} & ratio  & index & (mag)&(mag)\\
(1) & (2) & (3) & (4) & (5) & \multicolumn{3}{c}{(6)} & \multicolumn{3}{c}{(7)} & \multicolumn{3}{c}{(8)}  & (9) & (10)& (11)&(12)\\

\hline\hline
33   & 26\_D003 & 2018 & 51.2432 & -37.0096 & 16.9  & $\pm$ & 1.5  & -16.88 & $\pm$ & 0.03 & 20.94 & $\pm$ & 0.22 & 0.37 & 1.18 & 0.66  & -\\
37   & 25\_D241 & 2018 & 51.2893 & -36.3652 & 33.9  & $\pm$ & 4.3  & -17.37 & $\pm$ & 0.03 & 22.60 & $\pm$ & 0.31 & 0.68 & 1.09 & 0.45  & - \\
46   & 22\_D244 & 2018 & 51.6043 & -37.1278 & 8.5  & $\pm$ & 0.8  & -15.32 & $\pm$ & 0.03 & 21.60 & $\pm$ & 0.23 & 0.64 & 0.98 & 0.49  & -\\
100  & 16\_D417 & 2018 & 52.9485 & -35.0514 & 19.8  & $\pm$ & 2.5  & -16.14 & $\pm$ & 0.03 & 22.79 & $\pm$ & 0.32 & 0.76 & 1.46 & 0.71  & 1.31 \\
106  & 15\_D417 & 2018 & 53.1987 & -34.2387 & 10.6  & $\pm$ & 0.9  & -16.45 & $\pm$ & 0.03 & 20.65 & $\pm$ & 0.20 & 0.49 & 2.18 & 0.68  & 1.15 \\
113  & 15\_D107 & 2018 & 53.2794 & -34.8056 & 18.9  & $\pm$ & 2.3  & -15.74 & $\pm$ & 0.03 & 23.00 & $\pm$ & 0.30 & 0.69 & 1.19 & 0.48  & 0.86 \\
134  & 15\_D223 & 2016 & 53.5904 & -34.5925 & 6.5   & $\pm$ & 0.8  & -13.57 & $\pm$ & 0.03 & 22.64 & $\pm$ & 0.31 & 0.57 & 0.81 & 0.48  & 1.17 \\
135  & 15\_D384 & 2016 & 53.6284 & -34.2974 & 14.7  & $\pm$ & 1.7  & -15.96 & $\pm$ & 0.03 & 21.81 & $\pm$ & 0.28 & 0.47 & 1.58 & 0.64  & 1.19 \\
136  & 16\_D159 & 2016 & 53.6228 & -35.5465 & 17.5  & $\pm$ & 1.7  & -16.96 & $\pm$ & 0.03 & 21.84 & $\pm$ & 0.24 & 0.85 & 2.14 & 0.78  & 1.50 \\
143  & 16\_D002 & 2016 & 53.7467 & -35.1711 & 9.8   & $\pm$ & 0.6  & -17.83 & $\pm$ & 0.03 & 19.71 & $\pm$ & 0.14 & 0.85 & 4.34 & 0.74  & 1.76 \\
164  & 12\_D367 & 2016 & 54.0536 & -36.1665 & 9.9   & $\pm$ & 1.1  & -15.02 & $\pm$ & 0.03 & 22.07 & $\pm$ & 0.28 & 0.55 & 1.47 & 0.63  & 1.40 \\
178  & 10\_D302 & 2016 & 54.2027 & -34.2801 & 11.3  & $\pm$ & 1.7  & -14.18 & $\pm$ & 0.03 & 23.45 & $\pm$ & 0.38 & 0.71 & 1.24 & 0.58  & 1.28 \\
181  & 10\_D003 & 2016 & 54.2219 & -34.9384 & 9.6   & $\pm$ & 1.4  & -14.05 & $\pm$ & 0.03 & 23.07 & $\pm$ & 0.35 & 0.60 & 0.82 & 0.64  & 1.16 \\
182  & 11\_D279 & 2016 & 54.2263 & -35.3747 & 9.7   & $\pm$ & 0.6  & -17.11 & $\pm$ & 0.03 & 20.53 & $\pm$ & 0.17 & 0.96 & 2.43 & 0.81  & 1.42 \\
188  & 11\_D155 & 2015 & 54.2689 & -35.5901 & 12.2  & $\pm$ & 1.4  & -15.49 & $\pm$ & 0.03 & 22.65 & $\pm$ & 0.29 & 0.96 & 1.00 & 0.69  & 1.24 \\
195  & 10\_D014 & 2016 & 54.3472 & -34.9001 & 12.8  & $\pm$ & 1.9  & -14.46 & $\pm$ & 0.03 & 23.15 & $\pm$ & 0.37 & 0.54 & 1.06 & 0.66  & 1.24 \\
202  & 11\_D235 & 2015 & 54.5273 & -35.4399 & 13.3  & $\pm$ & 1.2  & -16.46 & $\pm$ & 0.03 & 21.34 & $\pm$ & 0.23 & 0.59 & 1.67 & 0.73  & 1.36 \\
203  & 10\_D189 & 2016 & 54.5382 & -34.5188 & 16.0  & $\pm$ & 1.9  & -16.03 & $\pm$ & 0.03 & 22.09 & $\pm$ & 0.29 & 0.55 & 1.46 & 0.66  & 1.50 \\
207  & 11\_D396 & 2015 & 54.5802 & -35.1291 & 9.6   & $\pm$ & 0.9  & -15.75 & $\pm$ & 0.03 & 21.72 & $\pm$ & 0.23 & 0.83 & 1.54 & 0.66  & 1.22 \\
211  & 11\_D339 & 2015 & 54.5895 & -35.2597 & 6.6   & $\pm$ & 0.6  & -15.29 & $\pm$ & 0.03 & 21.24 & $\pm$ & 0.21 & 0.75 & 1.66 & 0.64  & 1.22 \\
222  & 11\_D283 & 2015 & 54.8055 & -35.3714 & 16.1  & $\pm$ & 1.8  & -16.12 & $\pm$ & 0.03 & 22.53 & $\pm$ & 0.28 & 0.89 & 1.35 & 0.69  & 1.25 \\
223  & 11\_D079 & 2015 & 54.8321 & -35.7247 & 17.0  & $\pm$ & 2.5  & -15.06 & $\pm$ & 0.04 & 23.73 & $\pm$ & 0.36 & 0.89 & 1.00 & 0.59  & 1.28 \\
235  & 11\_D519 & 2016 & 55.0411 & -35.6291 & 42.3  & $\pm$ & 5.5  & -17.66 & $\pm$ & 0.03 & 22.79 & $\pm$ & 0.32 & 0.68 & 0.85 & 0.34  & 0.95 \\
245  & 11\_D458 & 2018 & 55.1410 & -35.0229 & 14.5  & $\pm$ & 1.8  & -15.60 & $\pm$ & 0.03 & 22.87 & $\pm$ & 0.30 & 0.92 & 1.51 & 0.62  & 1.37 \\
250  & 13\_D258 & 2018 & 55.1850 & -37.4083 & 9.2   & $\pm$ & 1.3  & -14.22 & $\pm$ & 0.03 & 23.06 & $\pm$ & 0.34 & 0.76 & 0.84 & 0.72  & 1.31 \\
252  & 11\_D069 & 2018 & 55.2100 & -35.7485 & 11.1  & $\pm$ & 1.2  & -15.71 & $\pm$ & 0.03 & 22.20 & $\pm$ & 0.26 & 0.94 & 1.21 & 0.69  & 1.60 \\
253  & 13\_D042 & 2018 & 55.2303 & -37.8376 & 10.9  & $\pm$ & 1.4  & -14.97 & $\pm$ & 0.03 & 22.45 & $\pm$ & 0.30 & 0.62 & 1.13 & 0.71  & 1.38 \\
263  & 5\_D000  & 2018 & 55.3856 & -34.8888 & 16.5  & $\pm$ & 1.4  & -17.50 & $\pm$ & 0.03 & 20.53 & $\pm$ & 0.20 & 0.48 & 1.37 & 0.48  & 0.98 \\
264  & 6\_D170  & 2015 & 55.3823 & -35.5896 & 10.3  & $\pm$ & 1.3  & -14.59 & $\pm$ & 0.03 & 22.23 & $\pm$ & 0.32 & 0.40 & 1.05 & 0.58  & 1.32 \\
266  & 6\_D455  & 2018 & 55.4222 & -35.1703 & 6.9   & $\pm$ & 0.6  & -15.47 & $\pm$ & 0.03 & 21.36 & $\pm$ & 0.21 & 0.89 & 1.17 & 0.66  & 1.45 \\
274  & 6\_D208  & 2015 & 55.5719 & -35.5407 & 12.0  & $\pm$ & 1.6  & -14.89 & $\pm$ & 0.03 & 23.22 & $\pm$ & 0.33 & 0.96 & 1.26 & 0.58  & 1.29 \\
277  & 6\_D002  & 2018 & 55.5949 & -35.1541 & 11.4  & $\pm$ & 0.7  & -18.39 & $\pm$ & 0.03 & 19.07 & $\pm$ & 0.15 & 0.58 & 1.89 & 0.65  & 1.64 \\
285  & 7\_D360  & 2018 & 55.7601 & -36.2734 & 32.6  & $\pm$ & 4.3  & -17.18 & $\pm$ & 0.03 & 22.82 & $\pm$ & 0.32 & 0.74 & 1.22 & 0.37  & 0.87 \\
298  & 6\_D098  & 2018 & 56.1851 & -35.6837 & 7.0   & $\pm$ & 0.7  & -14.64 & $\pm$ & 0.03 & 21.95 & $\pm$ & 0.25 & 0.71 & 1.19 & 0.62  & 1.43 \\
300  & 7\_D326  & 2018 & 56.2496 & -36.3198 & 20.8  & $\pm$ & 3.2  & -15.62 & $\pm$ & 0.03 & 23.37 & $\pm$ & 0.38 & 0.72 & 1.14 & 0.67  & 1.36 \\
301  & 7\_D000  & 2018 & 56.2649 & -35.9727 & 7.6   & $\pm$ & 0.4  & -17.46 & $\pm$ & 0.03 & 19.03 & $\pm$ & 0.13 & 0.54 & 2.12 & 0.72  & 1.64 \\
306  & 7\_D310  & 2018 & 56.4391 & -36.3461 & 7.4   & $\pm$ & 0.7  & -14.93 & $\pm$ & 0.03 & 21.56 & $\pm$ & 0.24 & 0.59 & 0.90 & 0.32  & 0.78 \\
B442 & 21\_D129 & 2018 & 51.7760 & -36.6368 & 4.4   & $\pm$ & 0.5  & -13.36 & $\pm$ & 0.03 & 22.24 & $\pm$ & 0.28 & 0.70 & 0.96 & 0.64  & 0.92 \\
B904 & 15\_D232 & 2016 & 53.4841 & -34.5618 & 5.1   & $\pm$ & 0.5  & -14.26 & $\pm$ & 0.03 & 21.80 & $\pm$ & 0.24 & 0.80 & 1.19 & 0.55  & 1.29\\                
\hline
\end{tabular}
\end{table*}

\section{Data \& Sample}
\label{sec:sample}

\begin{figure}
   \centering
   \includegraphics[width=8.3cm,keepaspectratio]{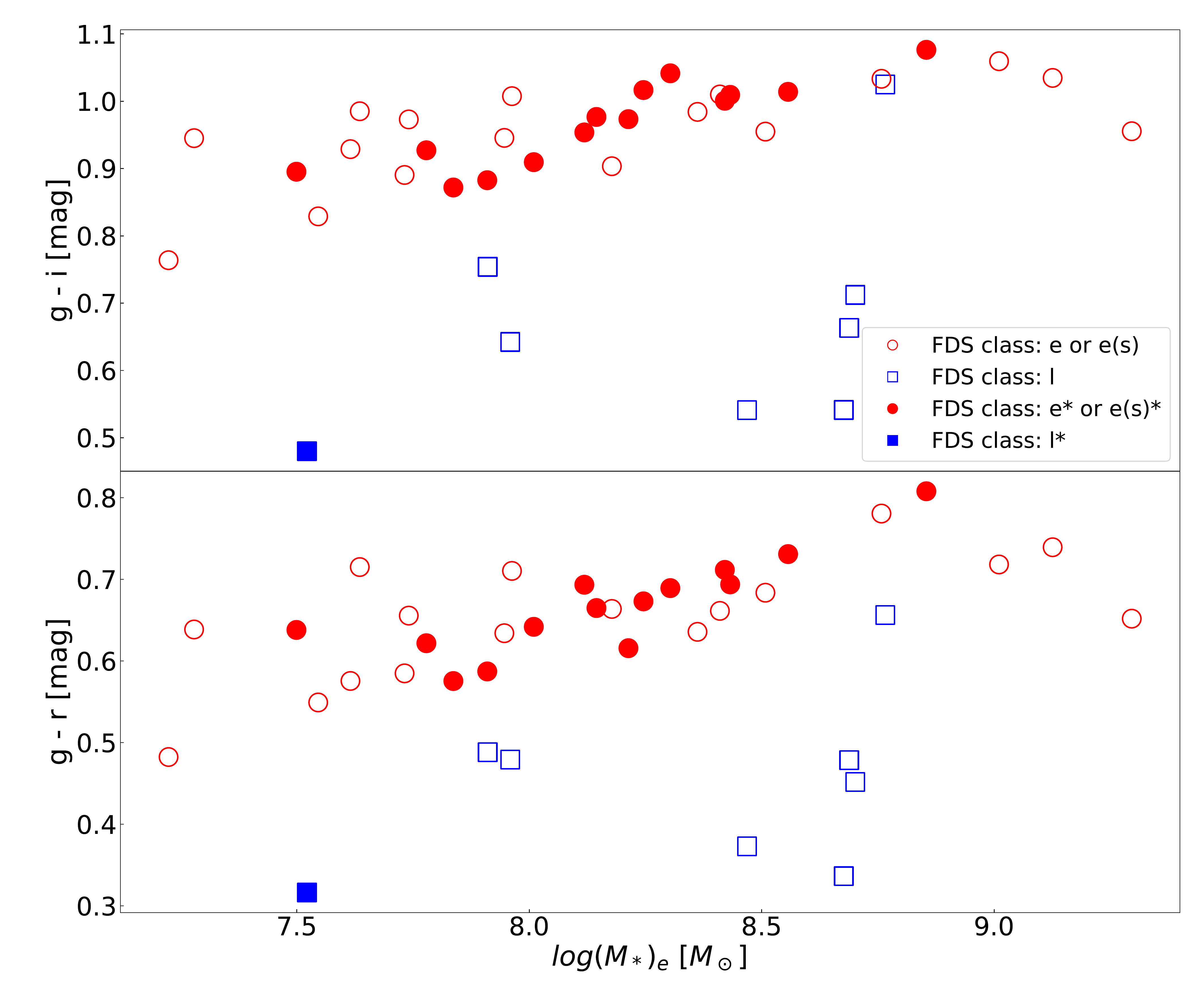}
         \caption{Color-mass relation for the DSAMI dwarf galaxies. The symbols indicate the morphological classification of the FDS, with early-type dwarfs indicated in red circles and irregular ones with blue squares. Galaxies with nuclear clusters are shown with filled symbols. One clearly sees that early-type dwarf galaxies form a red sequence, with the irregular dwarfs in the blue cloud below it.}
         \label{fig:CI-S}
\end{figure}

The SAMI--Fornax Dwarfs project started in 2015 with observations at the Anglo-Australian Telescope (AAT) using the Sydney-AAO Multi-Object Integral-field (SAMI) spectrograph to study the structure and origin of dwarf galaxies inside the Fornax cluster. SAMI is an integral-field spectrograph equipped with 13 fiber-based IFUs called hexabundles, and 26 pluggable sky fibers \citep{Bryant:2012,Bryant:2014,Bryant:2015}. With a field-of-view of $15''$ diameter, each hexabundle is made of 61 $1.6''$ diameter optical fibers. These hexabundles have a physical size $<$1 mm, with a filling fraction of 73 percent, and together with sky fibers, each fits into pre-drilled holes on a field plate. The plug plate, with a field-of-view of about 1 degree diameter, is installed at the AAT's Prime Focus Camera top end.
Studying the structure of low-mass galaxies, such as dwarf galaxies in clusters, requires high-resolution spectra and high S/N. For the SAMI--Fornax project, a 1500V grating in the blue and a 1000R in the red are used on the double-armed AAOmega spectrograph. This results in data with a resolution of approximately 1.05\AA{} (FWHM) in the blue ($\sim4660-5430${}\AA) and 1.6\AA{} in the red ($\sim6250-7350${}\AA), or an instrumental resolution of $\sigma\approx26.7 km/s$ (see Appendix A). As described in the Appendix, this allows us to measure velocity dispersions down to about 10 km/s. Given the low surface brightness of dwarf galaxies, we took fourteen ~30 minutes exposures for each field for a total exposure time of $\sim 7$\ hours. In this paper, we only describe the blue spectra, as these have the highest resolution and include the majority of the strong stellar absorption lines useful for kinematic analysis. A detailed description of the methods used to analyze the IFU spectra is given in the next section. Since SAMI can simultaneously observe 12 galaxies and one calibration star, the observing efficiency is significantly increased compared to single IFU instruments. Comparing to long-slit spectroscopy, integral field spectroscopy can usually observe a much larger fraction of the galaxy since it covers a 2D area rather than a long slit. The resolution of IFU spectrographs is much more stable and does not depend on aperture effects such as the seeing. Existing other IFU-surveys of galaxies (e.g., MANGA \citep{Bundy:2015}, ATLAS3D \citep{Cappellari:2011} , CALIFA \citep{Sanchez:2012}, SAMI \citep{Allen:2015}) mostly can not be used to study dwarf galaxies, since their spectral resolution is too low to determine the low velocity dispersions of dwarf galaxies. A possible exception is the work of \citet{Rys:2013, Rys:2014}, who studied the internal kinematics and stellar population in a sample of about ten dwarf galaxies with velocity dispersions down to $\sim$ 50 km/s, using very high S/N SAURON data. The same is the case for \citet{Penny:2016}, who studied a sample of galaxies with MANGA, down to the same velocity dispersion limit. There are, however, some long-slit spectroscopic studies with more than ten galaxies that are going down to a somewhat lower mass, such as \citet{Toloba:2014, Toloba:2015} and \citet{Penny:2015}.

In a total of 17 allocated nights in semesters 2015B, 2016B, and 2018B, we were able to observe the stellar kinematics for 38 dwarf galaxies ($M_r>-19$) and 16 giants. These giants will not be discussed in this paper. The observing strategy and a detailed description of the sample are given in paper 1. For completeness, a short description will also be given here. Our science targets were selected in 2015 from the Fornax Cluster Catalog (FCC, \citet{Ferguson:1989}), and in 2016 and 2018 from the Fornax Deep Survey (FDS, \citet{Venhola:2018}) dwarf galaxy catalog, as the much deeper Fornax Deep Survey was progressively completed. The primary survey targets are dEs with $-19<M_r<-14.5$ mag and effective surface brightness of $17<\mu_{r, e}<23$ mag~arcsec$^{-2}$. To allocate all 13 hexabundles of each pointing of the $1^o$ diameter plate, other dwarf galaxies, giant ellipticals, UCDs, and even background galaxies were also included in the target selection. The observations covered 90 objects in the central cluster up to $2^{\rm o}$  away from NGC 1399, and 9 objects in the Fornax A region. Approximately one-third of the objects were excluded, since their spectra were too faint and noisy for the measurement of kinematic parameters. We include 38 objects in this paper for which the stellar kinematics was accurate enough to include in scaling relations. Their photometric parameters, from the FDS, are given in Table 1. 

To be able to compare our data with the literature, we convert our magnitudes and surface brightnesses to the $V$-band using the transformation of \citet{Jordi:2006}:
\begin{figure}
   \centering
   \includegraphics[width=8.3cm,keepaspectratio]{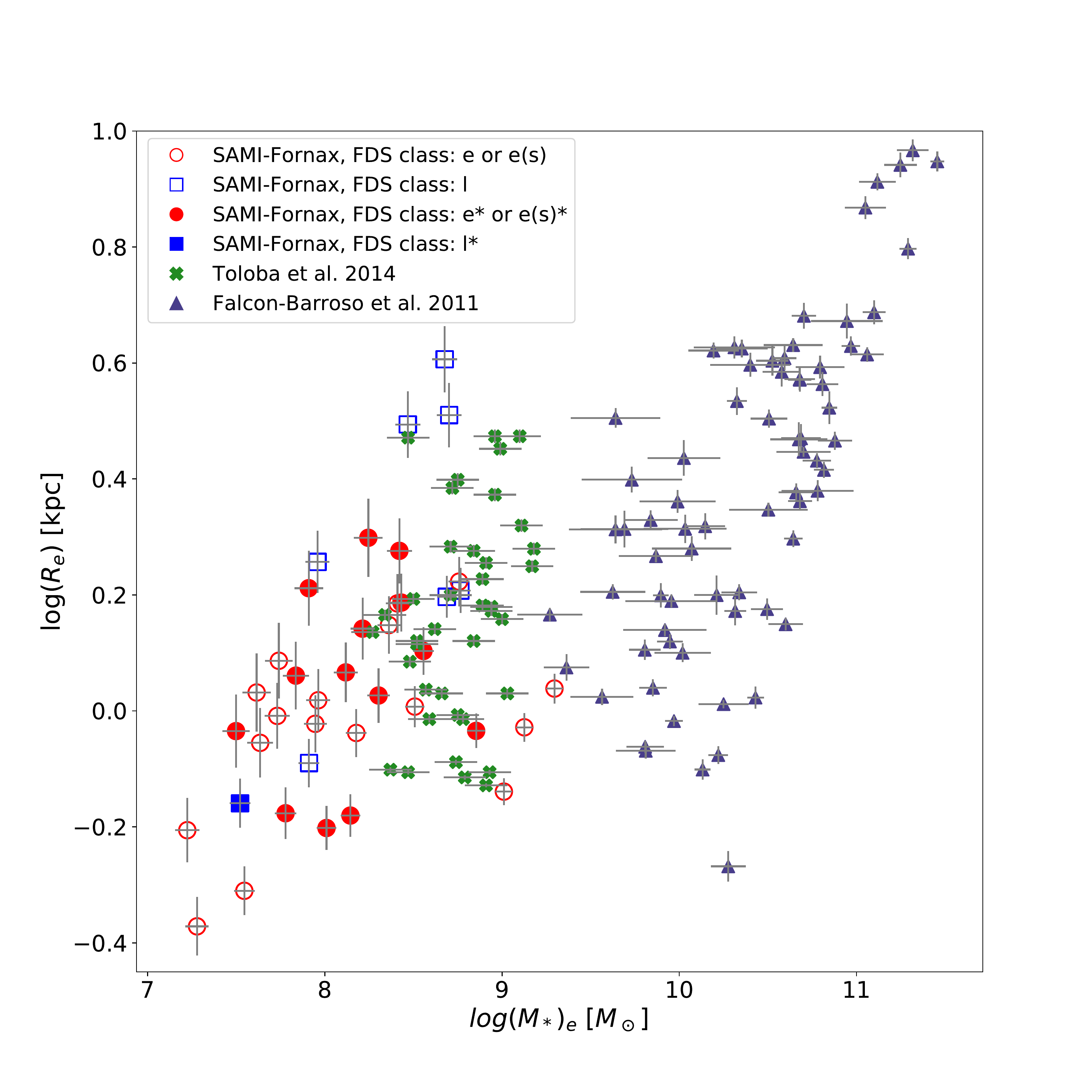}
         \caption{The relationship between effective radius and stellar mass for three spectroscopic data sets:  giant early-type galaxies (\citet{Falcon-Barroso:2011}, purple triangles), early type dwarf galaxies in the Virgo cluster (\citet{Toloba:2014}, green crosses) and our SAMI-Fornax dwarf galaxies. For the DSAMI sample the symbols indicate the morphological classification of the galaxies, from the FDS, with early-type dwarfs indicated in red and irregular ones with blue squares. Galaxies with nuclear clusters are shown with filled symbols.}
         \label{fig:params(R-M)}
\end{figure}

\begin{equation}
    V-g = (-0.565\pm0.001)(g-r) - (0.016\pm0.001)
\end{equation}

To obtain the mean effective surface brightness within R$_e$ we measured the total flux inside the ellipse with radius the $r$-band effective radius R$_e$ and axis ratio $q$, from  \citet{Venhola:2018}:

\begin{equation}
    \mu_{r,e} = m_{r,e} +2.5log(\pi~ R_e^2~q))
\end{equation}

We calculate the stellar mass within an effective radius using the formula of \citet{Taylor:2011}:

\begin{equation}
    log(M_*/M_{\odot})_e = 1.15 + 0.70(g-i)-0.4{M_{r,e}} +0.4(r-i)
\end{equation}

Here $M_{r,e}$ is the absolute magnitude inside an effective radius. 
In a future paper, we will revisit this issue when analyzing mass-to-light ratios determined from stellar population modelling on the spectra, together with optical and near-infrared colours. Here we will also investigate whether any IMF-dependent effects should be taken into account. Current studies, e.g., \citet{Mentz:2016}, indicate that this is probably not the case.

To illustrate the parameter space covered by the SAMI--Fornax data set, Fig.~\ref{fig:CI-S} shows the color-stellar mass relation for the $g-i$ and the $g-r$ color, measured within R$_e$. Comparing it with, e.g., Fig. 1 of \citet{Roediger:2017} one sees a similar scatter. Both samples include quiescent and star-forming dwarfs. Our sample contains a representative subsample of the low-mass Fornax dwarfs (for a detailed discussion, see paper 1).

In Fig.~\ref{fig:params(R-M)} we show the relation between stellar mass and effective radius for our sample, together with the literature samples of \citet{Toloba:2014} and \citet{Falcon-Barroso:2011}, samples which we will use throughout this paper. This figure can be compared to \citet{Kormendy:2009}, where this diagram is discussed in detail, to \citet{Misgeld:2011} or to \citet{Venhola:2019}, where a large, complete sample of dwarfs in the Fornax cluster is discussed. In all cases, a distinction is made between quiescent and star-forming objects. This diagram contains two sequences, one with radius decreasing with decreasing mass, and one with a more or less constant radius (the dwarfs).

\begin{figure*}
   \centering
   \includegraphics[width=17cm, keepaspectratio]{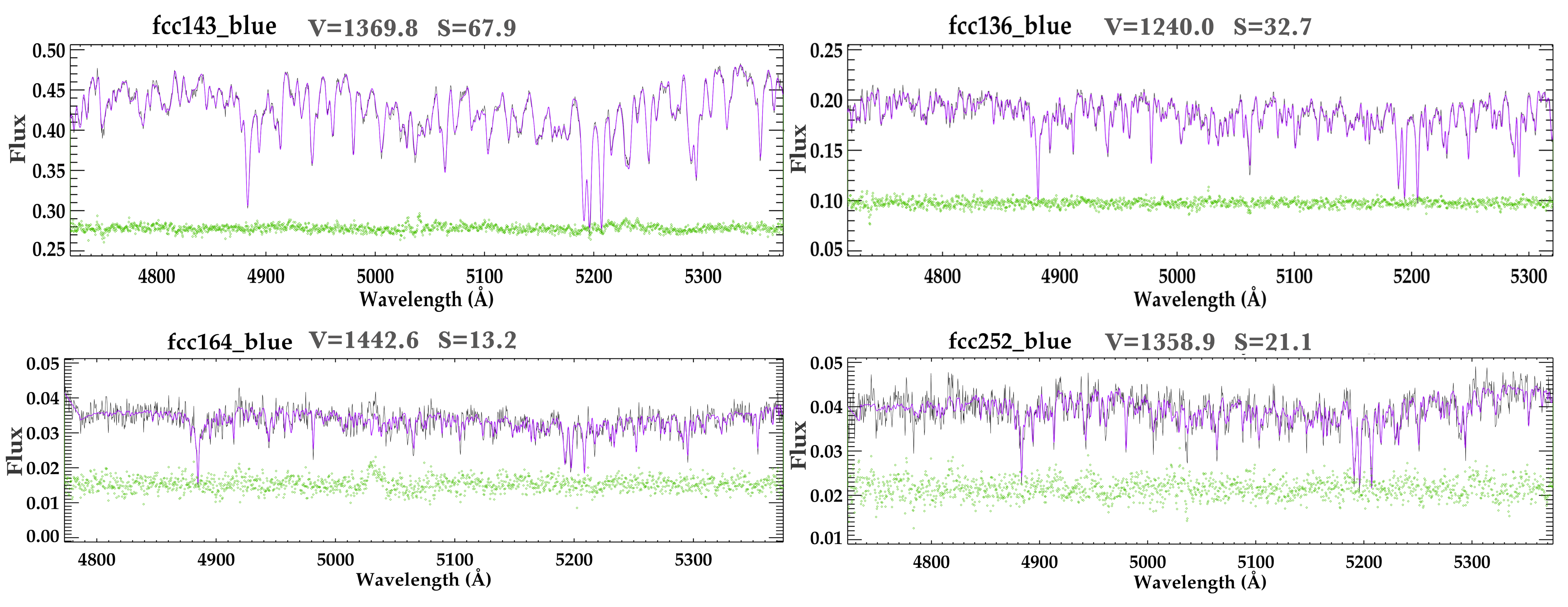}
  	 \caption{Four examples of pPXF fits to the SAMI-spectra of galaxies with a range in velocity dispersion. The observed spectrum is shown in black, while the derived best-fitted combination of stellar templates (from the Elodie stellar library, Prugniel \& Soubiran 2009) by pPXf is shown in purple. The residuals of the fit are shown with green dots, and are shifted upwards for visual clarity. The measured signal-to-noise, recovered mean velocity and velocity dispersion of each spectrum are indicated in each panel in units of km/s.}
         \label{fig:Best-Fits}
\end{figure*}

\section{Data Analysis}
We used the Penalized Pixel-Fitting (pPXF) routine of \citet{Cappellari:2004} to extract stellar kinematics from the absorption-line spectra of the galaxies. pPXF uses a maximum penalized likelihood approach in pixel space, intending to find a linear combination of observed stars that optimally fits the observed galaxy spectrum when convolved with a line-of-sight velocity distribution. In this process, we extract the first two Gauss-Hermite moments of the LOSVD (v, $\sigma$) of a spectrum integrated within a circular aperture.  No higher-order parameters were extracted, since the S/N ratio for the fainter galaxies does not allow this. We include an additive Legendre polynomial of the 6th degree to correct the stellar continuum shape for the optimized fitting.  For the spectral templates, the best match to our SAMI-Fornax wavelength range and high spectral resolution is the ELODIE \citep{Prugniel:2007} stellar library, 1959 spectra for 1503 stars with R$=10000$.
A few example-spectra of SAMI-Fornax galaxies and their corresponding fits are shown for the blue range in Fig.~\ref{fig:Best-Fits}. 
Some SAMI-Fornax galaxies were found to exhibit large emission lines, such as FCC33, FCC263, FCC181, for which the emission features were masked before extracting velocity  and velocity dispersion. We use the CLEAN keyword within pPXF, which applies an iterative sigma clipping to stabilize the fit against outliers, such as template mismatch \citep{Cappellari:2004}.

A minimum signal-to-noise must be available to get a reliable measurement of stellar kinematics. In Appendix A, we derive the instrumental resolution of our spectra $\sigma_{SAMI}=26.69\pm0.57(km/s)$ in the blue region by using standard stars observed during the survey. In Appendix B, we present a Monte-Carlo simulation to determine the uncertainties in the measured velocity dispersion as a function of $S/N$ ratio and resolution needed to interpret the data. In paper I, we spatially binned the data cubes using Voronoi binning \citep{Cappellari:2003}, so all bins had a minimum S/N ratio of 10\text. In this paper, to get single velocity dispersion values per galaxy, we integrate the spectra within an aperture of 15" diameter to get an approximation of $\sigma$ within R$_e$. For a typical dwarf in the SAMI-Fornax sample, the instrument's field-of-view covers about ${\rm R}_e/2$ area of the galaxy. For the fainter dwarfs, a larger fraction is covered.

\begin{figure}
   \centering
   \includegraphics[width=8.3cm,keepaspectratio]{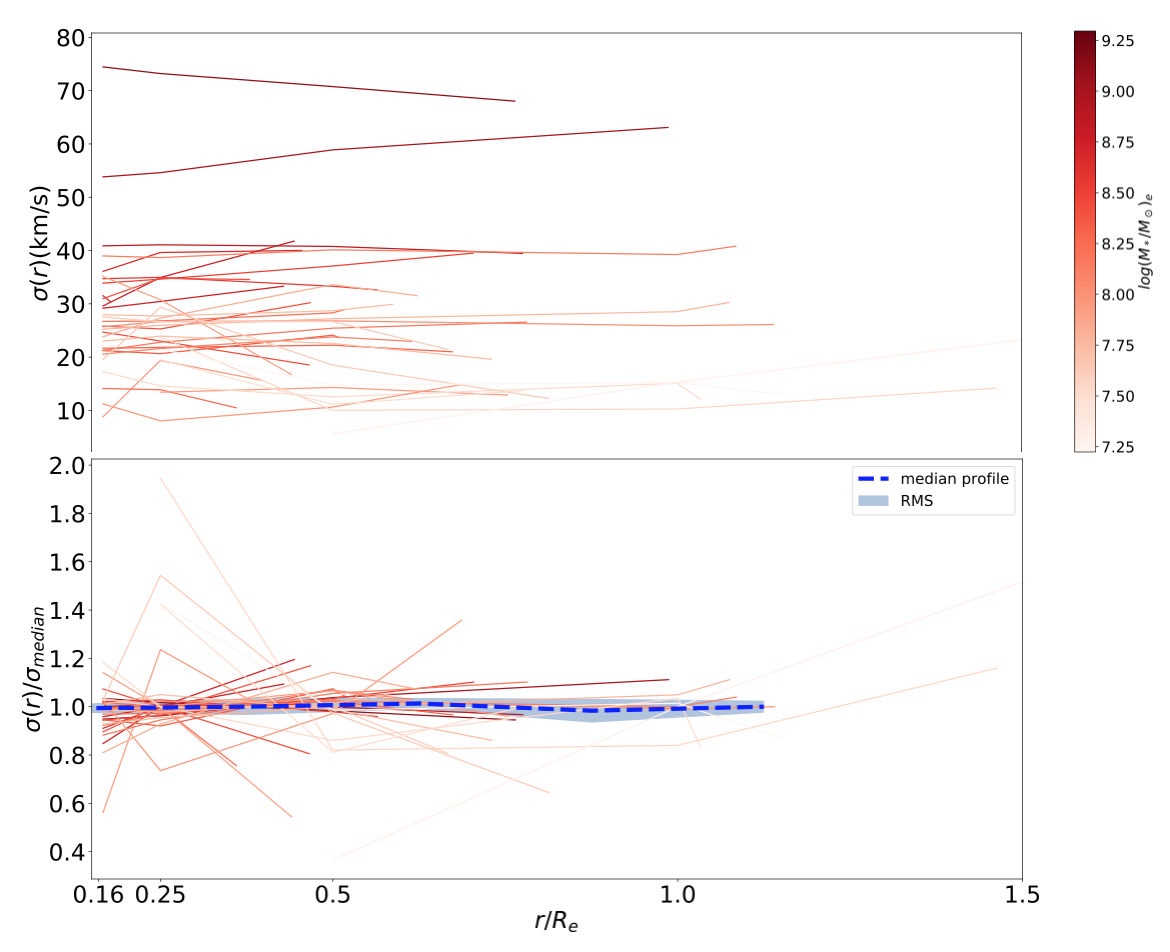}
         \caption{Integrated velocity dispersion profile and aperture corrections. The top panel shows the integrated velocity dispersion profile of the SAMI-Fornax dwarf galaxies connecting measurements inside $R_e/6$, $R_e/4$, $R_e/2$, R$_e$ and $r=7.5"$ (maximum of the SAMI FOV). The lines are color coded by stellar mass (eq. 3). In the lower panel each profile is normalized by its median value. The dashed black line is the total median profile of the individual normalized sigma profiles. The purple shadow indicates the standard deviation of the total median profile. As the radial profiles are remarkably flat, no aperture correction was applied.}
         \label{fig:aper-corr}
\end{figure}

In Fig.~\ref{fig:aper-corr}, we consider the velocity dispersion profiles of our dwarf SAMI-Fornax galaxies as a function of radius. The radial profiles of most of the galaxies are remarkably flat, apart from some galaxies, which only have a few points. (Fig. \ref{fig:aper-corr}, lower panel) presents the normalized velocity dispersion profiles and one median profile of all of them. These profiles show an almost flat behavior across different apertures. So to reach values within R$_e$, no aperture corrections were applied on the measured $\sigma_{7.5"}$. \\
A small number of 5 dEs and a bigger sample of 12 giant ellipticals in the Fornax3D survey overlap with our SAMI-Fornax primary and secondary samples, respectively. The  Fornax3D survey \citep{Iodice:2019} is observed with the MUSE instrument, an IFU with a field of view of 1'. When comparing the velocity dispersions of the two surveys, we see that by not aperture-correcting the velocity dispersions of our dwarf galaxies, we introduce an error of at most 10\%, while this is  $25\%$ for the giant galaxies, which are not used in this paper.

The uncertainties in the measurements of our radial velocities and velocity dispersions are estimated through Monte Carlo simulations, in which the input spectrum is perturbed by random values convolved to the resolution of the best-fitting template spectra. After 100 realizations, the mean and sigma of the distribution of each kinematic parameter are calculated. Uncertainties calculated in this way are presented in Appendix B. It is assumed that the uncertainties in luminosities and stellar masses are negligible compared to those in the velocity dispersions and the effective radii.

\section{Results from Scaling Relations}
Studies have shown that galaxy parameters sometimes follow tight 1D or 2D manifolds, which are ideal for studying their internal structure. Investigating the position of dEs on the FP and the relation between the perpendicular deviation from the plane with their other properties are some of the keys for better understanding the evolution of galaxies, their structure and stellar populations, or even their dark matter distribution \citep{Renzini:1993, Borriello:2003}. Here we concentrate mainly on the position of faint quiescent dwarfs with stellar masses between $10^{7}$ and $10^{8.5}$ M$_\odot$ on the FP. Note that these have barely been studied on the FP. Although the FP has been used mainly for quiescent galaxies, we also add our sample of star-forming dwarfs, intending to look for differences in structure between star-forming and quiescent dwarfs, a topic which has also barely been studied. We aim to study the dark matter content of objects, which are situated between brighter dwarfs, with relatively little dark matter, and dwarfs with stellar mass smaller than $10^{7}$ M$_\odot$, which, in the Local Group, have M/L ratios of up to 100 or more.

To study the FP extensively and check consistency with the literature, we extract the data of two older surveys. We compare our results with those of \citet{Falcon-Barroso:2011} (hereafter FB11), who studied ellipticals, lenticular, and spiral populations in the nearby Universe. Although quite a large sample of giants was observed as part of SAMI-Fornax, these observations cover only a small fraction of the galaxies, making them hard to use for the scaling relations. FB11 uses V-band photometry, so using the conversion shown in eq. 1, we convert our $g$-band photometry to V-band. This operation generally does not produce a scatter of more than 0.02 mag \citep{Jordi:2006}, which is much less than the scatter caused by the other parameters.
Moreover, we add to our analysis the dwarf ellipticals in the Virgo cluster by \citet{Toloba:2012} (hereafter T12), and a more extended sample of dEs by \citet{Toloba:2014} (hereafter T14), datasets similar in size and luminosity to our bright early-type dwarf galaxies.

\subsection{Fundamental Plane}
\label{FP}

\begin{figure*}
   \centering
   \includegraphics[width=15.4cm,keepaspectratio]{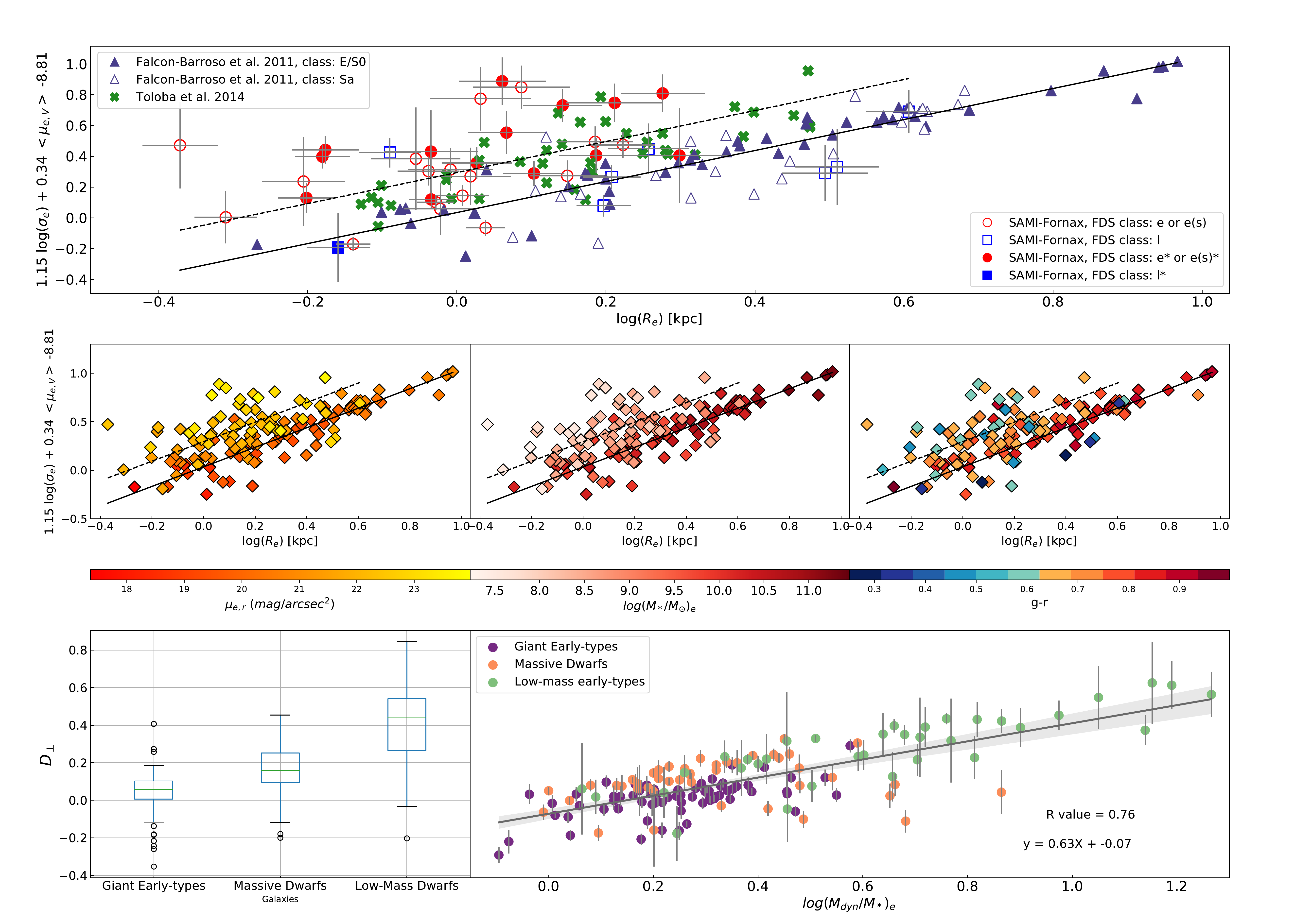}   	
  	 \caption{The FP defined by giant early-type galaxies. The solid line is the FP of \citet{Falcon-Barroso:2011}. The dwarf sample of T14 (green crosses) and the early-types of our SAMI sample (DSAMI(e)) are both situated on average above the FP of giants. These dwarfs are situated in a plane shifted by approximately 0.26 dex in the vertical direction above the FP.  Note, on the other hand, that star forming SAMI galaxies (DSAMI(l)) are positioned  close to and slightly below the FP of giants. In the top panel, the symbols are the same as in Fig.~\ref{fig:params(R-M)}. In the middle three panels the same figure is shown, with the points coloured based on resp. their effective surface brightness, stellar mass and $g-r$ colour. In the bottom panel the perpendicular displacement from the FP is shown as a function of dynamical mass, as calculated in Section 4.2. }
         \label{fig:FP-FB}
\end{figure*}

\begin{figure*}
   \centering
   \includegraphics[width=15.4cm,keepaspectratio]{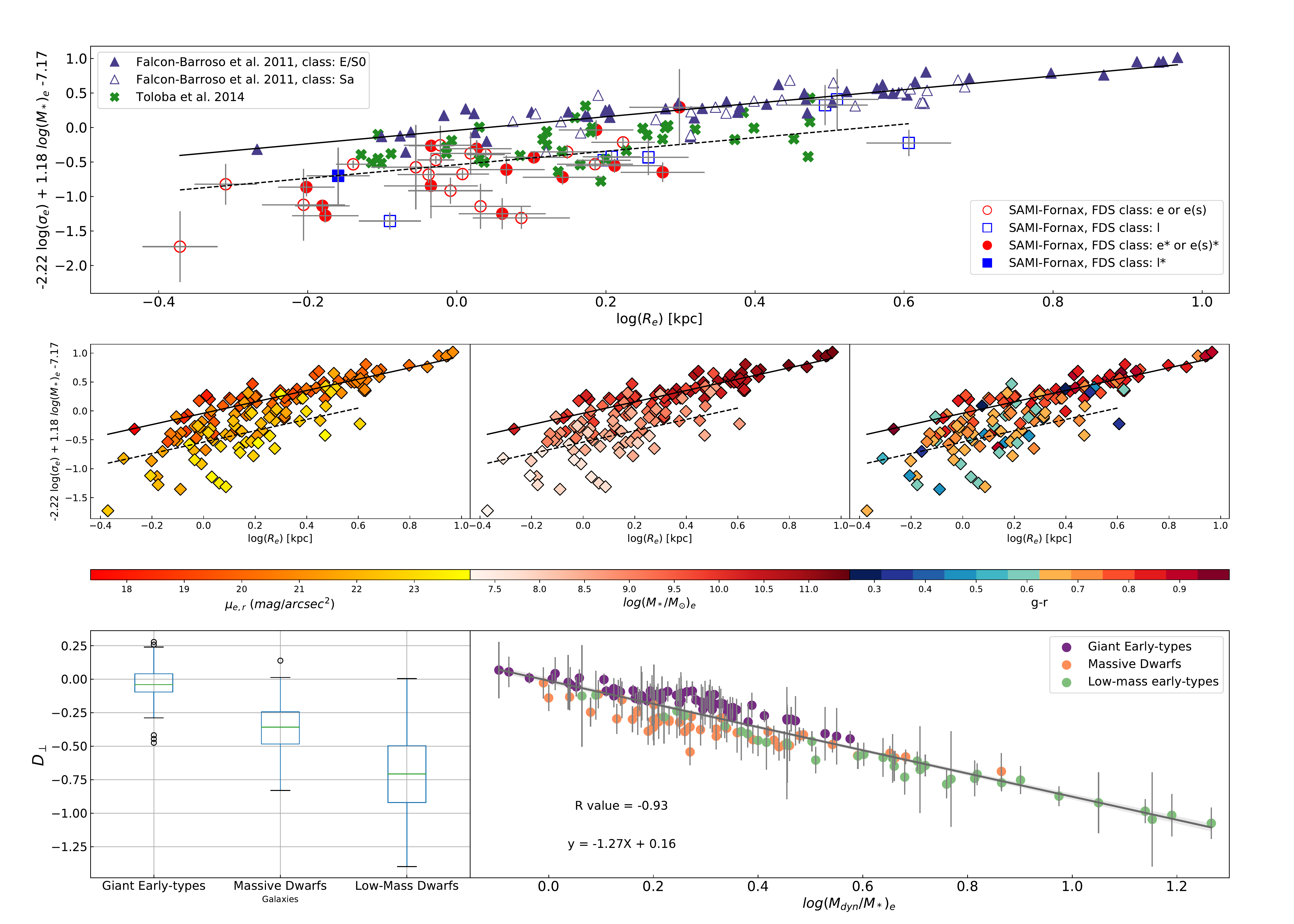}   	
  	 \caption{The Stellar Mass Fundamental Plane defined by giant early-type galaxies. For symbols see Figure \ref{fig:FP-FB}. In the top panel the solid line is made by using the LTS-planefit method of \citet{Cappellari:2013} and the FB11 dataset for Es and S0s, and the dashed line is vertically shifted about 0.47 dex below the $M_*$P, fitting the dwarf galaxies. In $M_*$P dIrrs move closer to the dEs and further away from the $M_*$P when compared to their position on the FP. In the middle three panels the same figure is shown, with the points coloured based on resp. their effective surface brightness, stellar mass and $g-r$ colour. In the bottom panel the perpendicular displacement from the M*P is shown as a function of dynamical mass, as calculated in Section 4.2.}
         \label{fig:MP-FB}
\end{figure*}

\begin{figure*}
   \centering
   \includegraphics[width=15.4cm,keepaspectratio]{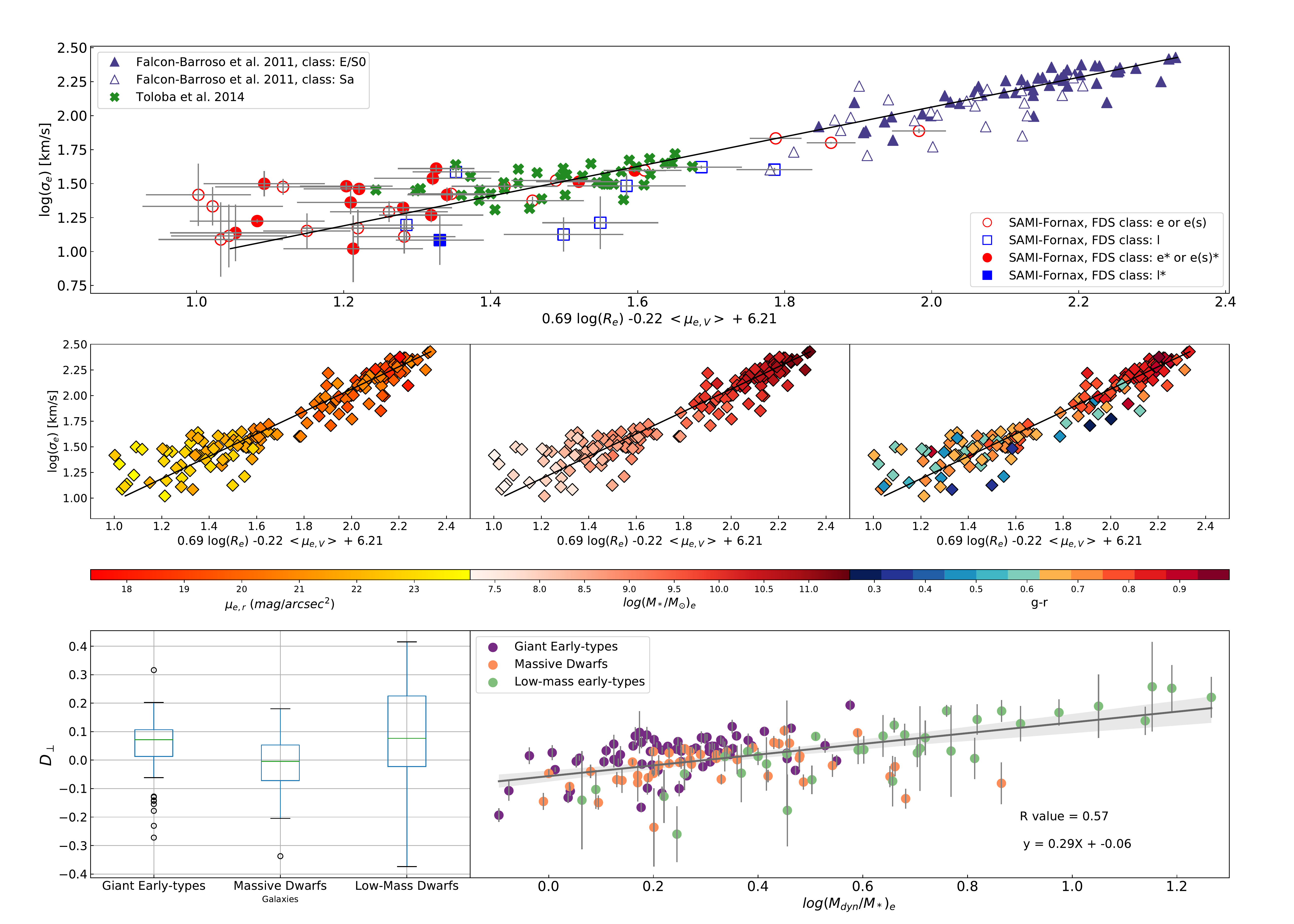}   	
  	 \caption{The FP defined by giant early-types and dwarf galaxies. The best-fit is made with the three data sets of this paper, T14 and F11 using the LTS-planefit
method of \citet{Cappellari:2013}.. For symbols see Figure 5. In the top figure, dwarfs are divided into early-type dwarfs from T14, shown with green crosses,
and SAMI-Fornax dwarfs which are shown with circles or squares depending on their morphological type. Giant early-types of FB11 are shown with filled
and empty triangles, representing of E/S0 or Sa galaxies respectively. In the middle three panels the same figure is shown, with the points coloured based on
resp. their effective surface brightness, stellar mass and $g$ -- $r$ colour. In the bottom panel the perpendicular displacement from the FP is shown as a function
of dynamical mass, as calculated in Section 4.2.}
         \label{fig:FP-LTS}
\end{figure*}

\begin{figure*}
   \centering
   \includegraphics[width=15.4cm,keepaspectratio]{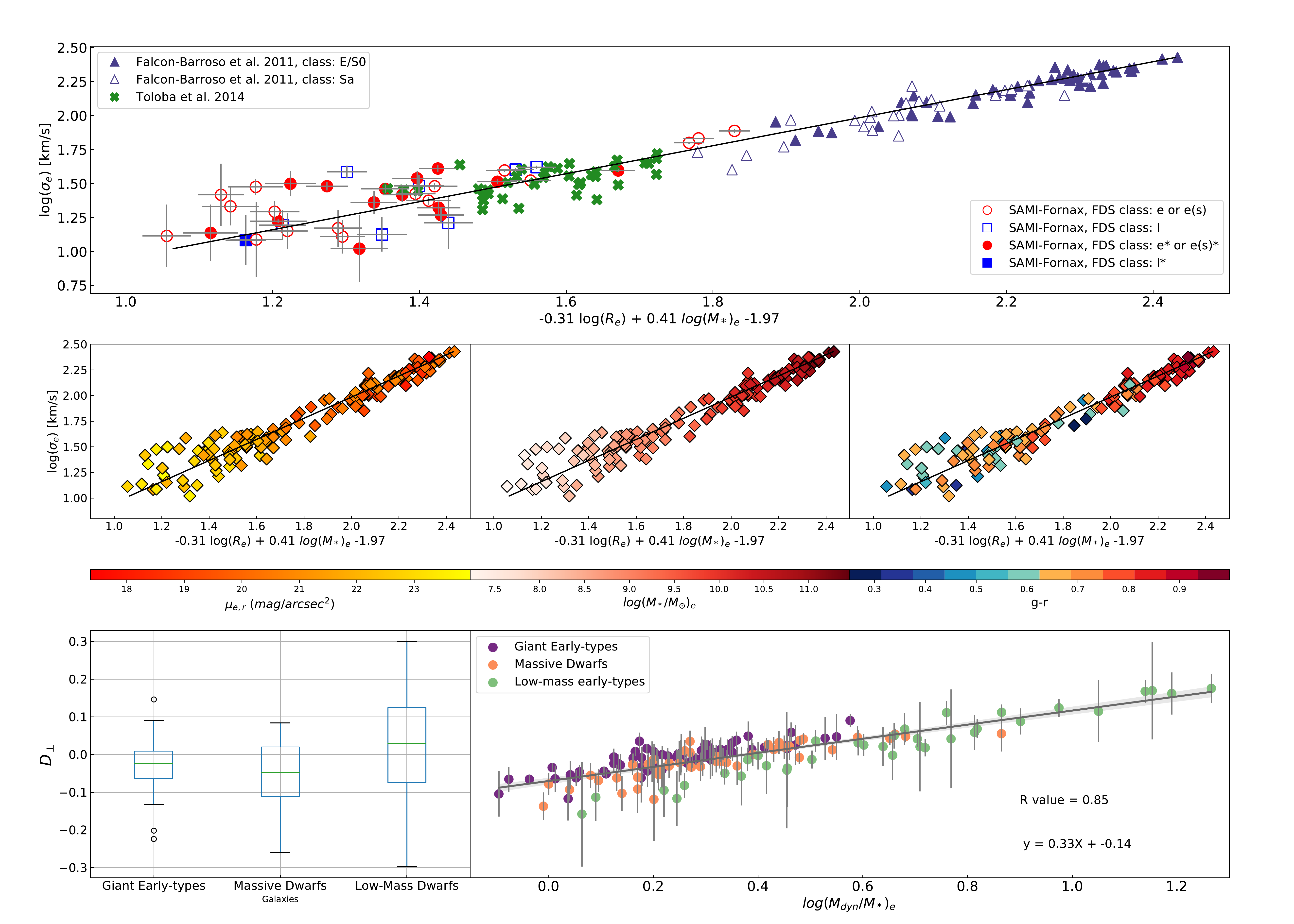}   	
  	 \caption{The Stellar Mass Fundamental Plane defined by giant early-type and dwarf galaxies. For symbols see Figure \ref{fig:FP-FB}. In the top panel the best-fitted line is made using the LTS-planefit method of \citet{Cappellari:2013} with the three samples. On the $M_*$P, the young dwarf irregular and blue giant spiral galaxies move closer to the plane and make a tighter plane. In the middle three panels the same figure is shown, with the points coloured based on resp. their effective surface brightness, stellar mass and $g-r$ colour. In the bottom panel the perpendicular displacement from the M*P is shown as a function of dynamical mass, as calculated in Section 4.2. }
         \label{fig:FP-Mass}
\end{figure*}

\begin{table}
\begin{center}
\caption{(Weighted) Perpendicular Averaged Deviations of early-type dwarf and irregular galaxies from the $log(R_e)$ projection of the FP and $M_*P$ defined by the giant early-type galaxies of FB11 (Fig.~\ref{fig:FP-FB} and Fig.~\ref{fig:MP-FB}).}
\begin{tabular}{cccc} %
\hline\hline\\ 
\textbf{Giants FP} & T14 (e) & DSAMI (e) & DSAMI (l) \\ [0.5ex] 
\hline\\
Weighted $\overline{D}_\bot$  & 0.231$\pm$0.049& 0.298$\pm$0.051 & 0.051$\pm$0.083\\[1ex] 
$\overline{D}_\bot$ & 0.242$\pm$0.042 & 0.365$\pm$0.047 & 0.040$\pm$0.079 
\\
\hline\\[0.5ex]
\textbf{Giants $M_*P$} & T14 (e) & DSAMI (e) & DSAMI (l) \\ [0.5ex] 
\hline\\
Weighted $\overline{D}_\bot $ & -0.376$\pm$0.077& -0.626$\pm$0.062 & -0.704$\pm$0.111\\[1ex]
$\overline{D}_\bot $ & -0.369$\pm$0.079 & -0.680$\pm$0.066 & -0.613$\pm$0.138\\
\hline
\end{tabular}
\label{table:d_rms_Giants_Planes}
\end{center}
The uncertainties in the measurements of the (weighted) perpendicular deviations are defined to be the scatter of the points around the mean value. Dwarf irregulars are found on the FP, so their scatter is measured w.r.t. FP itself (solid line). The scatter of early-type dwarfs on the FP and their scatter on the $M_*$P is measured w.r.t. the shifted planes (dashed line) above and below the FP and $M_*$P respectively.

\end{table}

Elliptical galaxies in the three-dimensional space of  half-light ratio radius, surface brightness, and internal velocity dispersion do not occupy the whole space but are confined to a surface, known as the FP. The empirical FP, which is a bivariate relation between R$_e$, I$_e$, and $\sigma_e$ is an indicator of galaxies being in virial equilibrium $R_e \propto \sigma^2 I_e^{-1} (M/L)^{-1}$ \citep{Brosche:1973,Binney:2008}. By assuming the mass-to-light ratio M/L of the galaxies to be a power-law function of $\sigma$ and $I_e$, the physical quantities can be replaced by observables, and the edge-on view of FP can be simplified to
\begin{eqnarray}
	\log(R_e) = A \log(\sigma) + B <\mu_e> ~+~ C
\end{eqnarray}
where $<\mu_e>$ is the mean effective surface brightness, and A, B and C are constants.
In the edge-on view of the FP, all distance-independent quantities, velocity dispersion, and surface brightness are collected on one side of the equation. The fact that R$_e$ and $I_e$ are strongly correlated is one of the keys to the FP's existence. In the face-on view we have
\begin{eqnarray}
\log(\sigma) = a \log(R_e) + b <\mu_e> ~+~ c
\end{eqnarray}
where these two correlated parameters are on one side, and the uncertainties are easier to analyse, since the observational uncertainties on both sides of the equation are not correlated. When calculating the error bars in this projection for $\sigma_e$ , we use the covariance matrix between R$_e$ and $\mu_e$, as taken from the FDS catalog \citep{Venhola:2018}.

The FP, the tightly correlated two-dimensional plane with low intrinsic scatter for giant galaxies \citep{Dressler:1987, Djorgovski:1987}, is derived by assuming that all galaxies are homologous, virialized systems with constant mass-to-light ratio.  Any deviation from these assumptions would result in changes in the plane's tilt and thickness. For example, changes in dynamical homology, the evolution of M/L as a function of stellar population (age, metallicity, and initial mass function), and possible dark matter content \citep{Graham:1997, DOnofrio:2006, Ciotti:1996, Busarello:1997, Trujillo:2004, Cappellari:2006} will change the constants A and B and results in variations in the thickness of the plane. The plane's thickness, i.e., the intrinsic scatter, has always been notably small and so a reliable distance indicator. In this 3D space, giant elliptical galaxies lie on the plane with a minor scatter of $\sim0.1$ dex \citep{Jorgensen:1996, Bernardi:2003}. 

\begin{table}
\caption{(Weighted) Perpendicular Averaged Deviations of dwarf galaxies from the $log(\sigma_e)$ projection of the FP and $M_*P$ defined over the full range of giant and dwarf galaxies (Fig.~\ref{fig:FP-LTS} and Fig.~\ref{fig:FP-Mass}).}
\begin{center}
\begin{tabular}{ccc} %
\hline\hline\\ 
\textbf{Giants+Dwarfs FP} & massive dwarfs & low-mass dwarfs \\ [0.5ex] 
\hline\\
Weighted $\overline{D}_\bot$ & -0.034$\pm$0.015 & 0.068$\pm$0.026\\[1ex]
$\overline{D}_\bot$ & -0.047$\pm$0.017 & 0.034$\pm$0.031\\[1ex]
\hline\\[0.5ex]
\textbf{Giants+Dwarfs $M_*$P} & massive dwarfs & low-mass dwarfs \\ [0.5ex] 
\hline\\
Weighted $\overline{D}_\bot$ & -0.010$\pm$0.013 & 0.078$\pm$0.025\\[1ex]
$\overline{D}_\bot$ & -0.019$\pm$0.014 & 0.046$\pm$0.027\\[1ex]
\hline
\end{tabular}
\end{center}
The uncertainties of the (weighted) deviations are calculated in the same way as for Table~\ref{table:d_rms_Giants_Planes}, and are defined to be the (weighted) scatter around the mean. Here dwarf galaxies are divided into two groups, of low-mass ($10^{7}<M_*\leq10^{8.5}M_{\odot}$) and massive dwarfs
($10^{8.5}<M_*\leq10^{9.5}M_{\odot}$). On the FP the two groups are situated on two opposite sides of the plane, with lower mass dwarfs deviating more in absolute terms.
\label{table:FaintBright-FPandMP}
\end{table}

There are multiple ways to fit a plane in the three-dimensional space ($logR_e$, $\mu_e$, $log\sigma$) and define its residuals.  In the direct method \citep{Bernardi:2003}, the residuals along one axis (mostly R$_e$) are minimized. In the orthogonal best-fitting method, the residuals in 3D are minimized. We investigated various orthogonal fit methods that have been used in the literature, such as the Weighted Least Square (WLS) method, the Weighted Least Absolute method (WLA), the improved fast Least Trimmed Squares method of \citet{Cappellari:2013}, and the Bayesian approach of 3D Gaussian Likelihood of \citet{Magoulas:2012}. It is important that before concluding the results and comparing different FP studies, we are sure that the choice of the best-fitting method does not influence the physical variations inferred from the different fits. In Appendix C, we show that the choice of fitting method does not affect our results (Table \ref{table:FP-DiffFits}), considering the data set used (DSAMI+T14/12+FB11). In FB11 the WLA method is used, which, compared to WLS, is less sensitive to a few outliers and more robust by treating all parameters symmetrically \citep{Jorgensen:1996, LaBarbera:2010}. Our choice in the regression method for the FP is the LTS-planefit of \citet{Cappellari:2013} as it provides the most stable approach. The uncertainties in each method's coefficients are derived through a bootstrap procedure or Monte Carlo simulations, for more details we refer to Appendix C.

Are the internal structure and kinematics of dwarf galaxies so similar to those of giant early-type galaxies that they fall on the same scaling relation? This question can be investigated by looking at how dwarfs are positioned on the FP defined by giant ellipticals. In Fig.~\ref{fig:FP-FB}, we present the edge-on view of the FB11 FP fit for giant early-type galaxies and show the location of dwarf galaxies relative to this plane. FB11 makes a fit (solid line) by minimizing the least absolute perpendicular scatter along the log$(R_e)$ axis. We find that the early-type dwarf galaxies in both the T14 and DSAMI samples are situated mostly above the FP of giant early-types, which is probably an indication that early-type dwarf galaxies have different internal structures and formation histories than giants. \\
In Table~\ref{table:d_rms_Giants_Planes}, we specify the (weighted) averaged perpendicular distance of dwarf galaxies of early-type dwarf (red circles) and late-type dwarf (blue squares) galaxies of SAMI-Fornax and early-type dwarfs of T14 individually. We see that early-type dwarfs within both samples deviate on average by 0.26 dex from the FP (in the log R$_e$ direction) and so are lying above the FP of massive hot stellar systems, indicated by a dashed line in Fig.~\ref{fig:FP-FB}. The scatter of the DSAMI(e) and T14 objects around the offset plane (dashed line) are not significantly different. The scatter around the relation does not seem to depend on the existence of a nucleus in the dwarf galaxies. Late-type dwarfs are situated on the average below the plane and close to the plane of giants. 

With the addition of the statistically large SAMI-Fornax sample, we can now determine an accurate FP fit to both giant and dwarf galaxies. 
In Fig.~\ref{fig:FP-LTS} we make a best-fit by minimizing the perpendicular scatter along with the $log(\sigma)$ axis, and so we have FP plots in which log($\sigma$) lies on one axis and the combination of
the other two parameters ($log(R_e)$ and $<\mu_e>$) on the other axis. In this way, the dependent parameters (R$_e$ and $\mu_e$) are taken together, making the interpretation easier. We find:

\begin{align}
\log(\sigma_e)  & =  0.693\pm0.033 ~ \log(R_e)\nonumber \\
 & +   (-0.218\pm0.007) <\mu_{e,V}>\nonumber \\
 &	+  6.21\pm0.15\nonumber \\
\end{align}

As expected, the FP (Fig.~\ref{fig:FP-LTS}) has a lower scatter than the FJ (Fig.~\ref{fig:FJ}) relation, since one more parameter is included in the fit. We do not see any pattern with nucleation in the dwarfs. \\
In Table \ref{table:FaintBright-FPandMP}, we divide the dwarf galaxies (DSAMI+T14) into low-mass ($10^{7}<M_*\leq10^{8.5}M_{\odot}$) and higher mass ($10^{8.5}<M_*\leq10^{9.5}M_{\odot}$) ones. Since our FP is derived using the weighted LTS method, we measure the perpendicular weighted averaged distance from the plane and their weighted RMS. The non-weighted values are included as well for consistency with previous works. We see that, as we go towards lower mass objects, their deviation from the plane becomes higher. Moreover, the two groups of dwarfs are located on opposite sides of the plane. This effect can be due to some small differences between their stellar populations, or their DM fraction. In the next paragraphs, we will investigate the contribution of each parameter and will correct the effects of stellar populations to clearly assess the differences in the internal structure (e.g. dark matter content) between early-type dwarf galaxies and giant early-type galaxies.

\subsection{Stellar Mass (Fundamental) Plane}
\label{StMass}
\begin{figure}
   \centering
   \includegraphics[width=8cm,keepaspectratio]{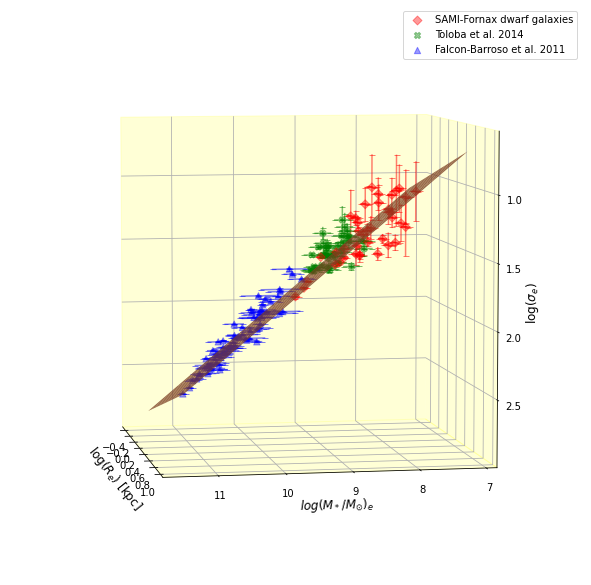}   	
  	 \caption{The Stellar Mass Fundamental Plane defined by giant early-type and dwarf galaxies in a 3D representation. For details, see Fig.~\ref{fig:FP-Mass}}
         \label{fig:FP-Mass-3D}
\end{figure}

When observing the position of dwarf galaxies w.r.t. the FP (Fig.~\ref{fig:FP-LTS}), we see that most objects below the plane are blue dwarfs, often classified as irregulars (classified as l or l* in \citet{Venhola:2018}), and some of the giant spiral galaxies. Also, in the FP defined by giant Es and S0s (Fig.~\ref{fig:FP-FB}), we saw that dIrrs are positioned quite differently compared to dEs in the T14 and DSAMI samples.
To remove the influence of different stellar populations, we use a new form of the FP in which luminosity is replaced by stellar mass, or, in practice, stellar mass replaces surface brightness. By using the LTS-planefit method of \citet{Cappellari:2013}, we derive a stellar-mass plane defined by giant early-types of FB11 and one defined by the whole range of galaxies from giant early-types to dwarf galaxies, which can be seen in Fig.~\ref{fig:MP-FB} and Fig.~\ref{fig:FP-Mass}, respectively. We abbreviate this plane as the M$_*$P. A 3d representation of Fig.~\ref{fig:FP-Mass} is shown in Fig.~\ref{fig:FP-Mass-3D}.\\ Our best fit to the Stellar Mass Fundamental Plane, defined by the FB11 sample of giants is:

\begin{align}
    \log(R_e) & = (-2.22\pm0.18)~ \log(\sigma_e)\nonumber \\
    & + (1.183\pm0.057)~ \log(M_*)\nonumber \\
    & + -7.17\pm0.33\nonumber \\
    \end{align}

\noindent The Stellar Mass Fundamental Plane defined by all galaxies:

\begin{align}
    \log(\sigma_e) & = (-0.307\pm0.027)~ \log(R_e)\nonumber \\
    & + (0.407\pm0.007) \log(M_*)\nonumber \\
    & + (-1.971\pm0.070)\nonumber \\
\end{align}

Correspondingly in Table \ref{table:FaintBright-FPandMP}, the perpendicular weighted averaged deviation and scatter of the points around the planes are measured. It is seen that the perpendicular deviation of dwarfs decreases in the $M_*$P, making the Stellar Mass Fundamental Plane thinner than the classic FP. However, the  lower mass dwarfs show a clear offset with respect to higher mass dwarfs and giants.
In the standard LCDM cosmology, the deviation of dwarfs from the $M_*$P can be explained by an increased dark matter fraction in them \citep{Zaritsky:2006, Zaritsky:2008, Graves:2010, Falcon-Barroso:2011, Toloba:2012}. Alternatively, this may also be due to tidal effects (i.e., galaxies are out of virial equilibrium), or it would also be expected in modified Newtonian dynamics, MOND \citep{Dabringhausen:2016}, without the need for dark matter. In \citet{Venhola:2021} we show that in Fornax there are clear signs that in the center of the cluster the surface brightness of dwarfs is lower than further out as a result of tidal forces,indicating that indeed many of the dwarf galaxies in the center of the cluster can be out of equilibrium.

\subsection{Dynamical Mass}
\label{DynMass}
How M/L varies with other photometric and kinematic parameters of the galaxies, especially with $\sigma_e$, what the scatter is (\citealp{Reda:2005}), and the dependence on mass, is an important goal of this work \citep{Cappellari:2006, Graves:2010}.  Here we measure the dynamical mass and the dark matter fraction of our galaxies in a different way, following the formalism of \citet{Wolf:2010} derived from the spherical Jeans equation. We use a straightforward analysis to determine the dynamical mass, which is simple and practical, without making too many assumptions. The virial mass estimator we use is robust against velocity anisotropy when computed within the half-light radius, where models with different anisotropy values cross each other (see also e.g., Figure 1 of \citet{Wolf:2010}.
Comparison with Local Group galaxies suffers from the fact that we only have 2d projected radial properties. However, \citet{Wolf:2010}, using the spherical Jeans equation, showed that within the $r_3$ radius, the difference between the projected dispersion and the dispersion from individual stars is insignificant. $r_3$ is the radius at which the log-slope of the 3D tracer density profile is -3, and for dispersion supported galaxies, it is close to the 3D half-light radius $R_{1/2}$. So the dynamical mass of dispersion supported galaxies can be defined as:
\begin{eqnarray}
	M_{1/2}\simeq 0.93 (\frac{{\sigma_{los}}^2}{{km}^2s^{-2}})(\frac{R_e}{kpc})~ 10^6~M_{\odot}
\end{eqnarray}
where R$_e$ is the 2D projected half-light radius and $\sigma_{los}$ is the line-of-sight velocity dispersion. Given the almost constant velocity dispersion, we use $\sigma_e$ for $\sigma_{los}$.

\begin{figure}
   \centering
   \includegraphics[width=8cm,keepaspectratio]{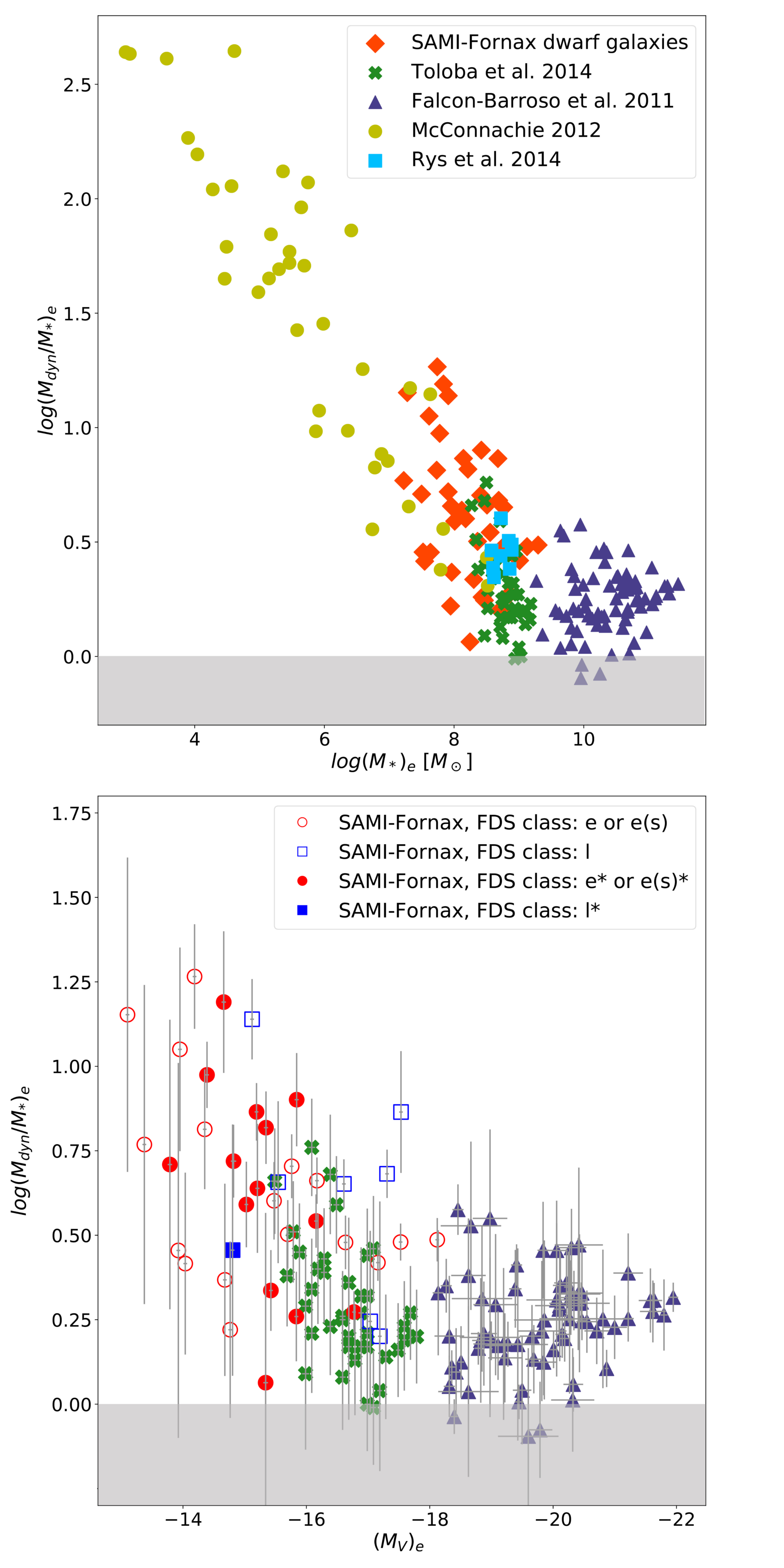}   		
  	 \caption{Effective dynamical to stellar mass ratios vs. stellar mass - In the top panel the three datasets discussed in this paper are shown together with the Local Group sample of \citet{McConnachie:2012} and the dwarf sample of \citet{Rys:2014}. The lower figure is a zoom-in of the upper figure including error bars, in which the symbols are the same as in the previous plots. In grey is indicated the prohibited area, where the dynamical M/L ratio is lower than the stellar one. }
         \label{fig:ML}
\end{figure}

In Fig.~\ref{fig:ML} we plot the ratio of dynamical and stellar mass for the galaxies in the three samples of DSAMI, T14, and FB11 in addition to the dwarf galaxies in the Local Group from \citet{McConnachie:2012} (hereafter MC12). MC12 study different aspects of the nearby population of Local Group dwarf galaxies. Their sample contains satellite systems, quasi-isolated members of the Local Group, and nearby isolated dwarf galaxies. The figure shows the classical minimum in DM ratio at about M$_*$ = $10^{9.5} M_\odot$, corresponding to a halo mass of $10^{11.5} M_\odot$, which is thought to be the halo mass with the highest stellar mass fraction \citep{Silk:2012}. In the mass region of overlap between our sample and the Local Group, the inferred dark matter fractions are indistinguishable between both environments. Error bars are given in the bottom part of Fig.~\ref{fig:ML} and in Table \ref{table:MLRatios}. In Table \ref{table:M_Dy/M_St}, we bin separately the ML ratios of dwarfs with lower stellar mass ($10^{7.0} < M_*/M_\odot \leq 10^{8.5}$), and higher values ($10^{8.5} < M_*/M_\odot \leq 10^{9.5}$) and find that the ML ratios are on the average considerably larger than for the more massive galaxies. 

\begin{table}
\centering
\caption{Weighted Averaged ratio of dynamical and stellar mass  of low-mass and more massive dwarf galaxies}
\begin{tabular}{ccc} %
\hline\hline\\ 
\textbf{$log(M_{dyn}/M_*)_e$} & massive dwarfs & low-mass dwarfs \\ [0.5ex] 
\hline\\
Weighted Average & 0.410$\pm$0.030 & 0.692$\pm$0.046\\[1ex]
Average & 0.328$\pm$0.030 & 0.635$\pm$0.054\\[1ex]
\hline
\end{tabular}
\label{table:M_Dy/M_St}
\end{table}

\begin{table*}
\begin{center}
\caption{Kinematic and mass measurements of the SAMI -- Fornax Dwarf Survey}
\begin{tabular}{lllllllllllllll}
\hline
FCC  & FDS  & \multicolumn{3}{c}{$V_{los}$} & \multicolumn{3}{c}{$\sigma_e$} & $\lambda_e$ & \multicolumn{3}{c}{$log(M_*)_e$} & \multicolumn{3}{c}{$log(M_{dyn})_e$}  \\
ID  & Class. & \multicolumn{3}{c}{(km/s)} & \multicolumn{3}{c}{(km/s)} &  & \multicolumn{3}{c}{$(M_\odot)$} & \multicolumn{3}{c}{$(M_\odot)$}  \\
(1)  & (2)  & \multicolumn{3}{c}{(3)} & \multicolumn{3}{c}{(4)} & (5) & \multicolumn{3}{c}{(6)} & \multicolumn{3}{c}{(7)}  \\
\hline\hline

33   & l      & 1985 & $\pm$ & 1 & 42  & $\pm$& 1  & 0.46  & 8.77  & $\pm$ & 0.06 & 9.42  & $\pm$ & 0.04  \\
37   & l      & 1834 & $\pm$ & 4 & 16  & $\pm$& 7  & - & 8.70  & $\pm$ & 0.07 & 8.90  & $\pm$ & 0.39  \\
46   & l      & 2267 & $\pm$ & 4 & 39  & $\pm$& 4  & 0.38  & 7.91  & $\pm$ & 0.06 & 9.05    & $\pm$ & 0.10  \\
100  & e*     & 1632 & $\pm$ & 3 & 35  & $\pm$& 4  & 0.28  & 8.42  & $\pm$ & 0.07 & 9.32  & $\pm$ & 0.12  \\
106  & e(s)   & 205 & $\pm$ & 1 & 40  & $\pm$& 1  & 0.14  & 8.51  & $\pm$ & 0.05 & 9.17  & $\pm$ & 0.04  \\
113  & l      & 138 & $\pm$ & 2 & 16  & $\pm$& 4  & - & 7.96  & $\pm$ & 0.07 & 8.61  & $\pm$ & 0.23  \\
134  & e      & 1404 & $\pm$ & 3 & 13  & $\pm$& 7  & - & 7.22  & $\pm$ & 0.07 & 7.99  & $\pm$ & 0.47  \\
135  & e(s)   & 1252 & $\pm$ & 1 & 24  & $\pm$& 2  & 0.31  & 8.36  & $\pm$ & 0.06 & 8.86  & $\pm$ & 0.08  \\
136  & e      & 1240 & $\pm$ & 1 & 33  & $\pm$& 1  & 0.15  & 8.76  & $\pm$ & 0.06 & 9.23  & $\pm$ & 0.05  \\
143  & e(s)   & 1370 & $\pm$ & 1 & 68  & $\pm$& 1  & 0.15  & 9.13  & $\pm$ & 0.05 & 9.60  & $\pm$ & 0.03  \\
164  & e(s)   & 1443 & $\pm$ & 1 & 13  & $\pm$& 4  & 0.08  & 7.95  & $\pm$ & 0.06 & 8.16  & $\pm$ & 0.25  \\
178  & e      & 1273 & $\pm$ & 4 & 22  & $\pm$& 7  & - & 7.62  & $\pm$ & 0.08 & 8.66  & $\pm$ & 0.29  \\
181  & e*     & 1215 & $\pm$ & 2 & 14  & $\pm$& 7  & - & 7.50  & $\pm$ & 0.08 & 8.21  & $\pm$ & 0.42  \\
182  & e(s)*  & 1695 & $\pm$ & 1 & 39  & $\pm$& 1  & 0.18  & 8.85  & $\pm$ & 0.05 & 9.13  & $\pm$ & 0.03  \\
188  & e*     & 1088 & $\pm$ & 3 & 23  & $\pm$& 5  & 0.20  & 8.12  & $\pm$ & 0.07 & 8.76  & $\pm$ & 0.18  \\
195  & e      & 1295 & $\pm$ & 3 & 30  & $\pm$& 4  & - & 7.74  & $\pm$ & 0.08 & 9.01  & $\pm$ & 0.13  \\
202  & e*     & 850  & $\pm$ & 1 & 33  & $\pm$& 1  & 0.13  & 8.56  & $\pm$ & 0.06 & 9.10  & $\pm$ & 0.05  \\
203  & e(s)   & 1143 & $\pm$ & 1 & 30  & $\pm$& 2  & 0.33  & 8.41  & $\pm$ & 0.07 & 9.11  & $\pm$ & 0.07  \\
207  & e      & 1428 & $\pm$ & 2 & 27  & $\pm$& 3  & 0.31  & 8.18  & $\pm$ & 0.06 & 8.78  & $\pm$ & 0.10  \\
211  & e*     & 2295 & $\pm$ & 2 & 26  & $\pm$& 3  & 0.11  & 8.01  & $\pm$ & 0.06 & 8.60  & $\pm$ & 0.11  \\
222  & e*     & 809  & $\pm$ & 1 & 19  & $\pm$& 2  & 0.32  & 8.43  & $\pm$ & 0.07 & 8.69  & $\pm$ & 0.11  \\
223  & e*     & 702  & $\pm$ & 9 & 17  & $\pm$& 1  & - & 7.91  & $\pm$ & 0.08 & 8.63  & $\pm$ & 0.07  \\
235  & l      & 1997 & $\pm$ & 5 & 30  & $\pm$& 5  & - & 8.68  & $\pm$ & 0.07 & 9.54  & $\pm$ & 0.16  \\
245  & e*     & 2191 & $\pm$ & 2 & 29  & $\pm$& 2  & 0.19  & 8.21  & $\pm$ & 0.07 & 9.03  & $\pm$ & 0.08  \\
250  & e      & 1261 & $\pm$ & 4 & 12  & $\pm$& 8  & - & 7.64  & $\pm$ & 0.07 & 8.09  & $\pm$ & 0.55  \\
252  & e*     & 1359 & $\pm$ & 1 & 21  & $\pm$& 2  & 0.15  & 8.30  & $\pm$ & 0.06 & 8.64  & $\pm$ & 0.10  \\
253  & e      & 1645 & $\pm$ & 2 & 15  & $\pm$& 5  & 0.26  & 7.96  & $\pm$ & 0.07 & 8.33  & $\pm$ & 0.27  \\
263  & l      & 1742 & $\pm$ & 1 & 40  & $\pm$& 1  & 0.15  & 8.69  & $\pm$ & 0.06 & 9.37    & $\pm$ & 0.04  \\
264  & e      & 1923 & $\pm$ & 3 & 20  & $\pm$& 3  & - & 7.73  & $\pm$ & 0.07 & 8.54  & $\pm$ & 0.16  \\
266  & e*     & 1567 & $\pm$ & 2 & 41  & $\pm$& 3  & 0.17  & 8.14  & $\pm$ & 0.06 & 9.01  & $\pm$ & 0.06  \\
274  & e*     & 972  & $\pm$ & 5 & 32  & $\pm$& 7  & - & 7.84  & $\pm$ & 0.07 & 9.03  & $\pm$ & 0.19  \\
277  & e(s)   & 1645 & $\pm$ & 2 & 77  & $\pm$& 3  & 0.27  & 9.30  & $\pm$ & 0.05 & 9.78  & $\pm$ & 0.04  \\
285  & l      & 893  & $\pm$ & 2 & 13  & $\pm$& 4  & - & 8.47  & $\pm$ & 0.07 & 8.71  & $\pm$ & 0.26  \\
298  & e*     & 1656 & $\pm$ & 2 & 30  & $\pm$& 2  & 0.18  & 7.78  & $\pm$ & 0.06 & 8.75  & $\pm$ & 0.07  \\
300  & e*     & 1400 & $\pm$ & 2 & 11  & $\pm$& 6  & 0.28  & 8.25  & $\pm$ & 0.08 & 8.31  & $\pm$ & 0.49  \\
301  & e(s)   & 1047 & $\pm$ & 1 & 63  & $\pm$& 1  & 0.39  & 9.01  & $\pm$ & 0.05 & 9.43  & $\pm$ & 0.03  \\
306  & l*     & 894  & $\pm$ & 2 & 12  & $\pm$& 5  & 0.52  & 7.52  & $\pm$ & 0.06 & 7.98  & $\pm$ & 0.37  \\
B442 & e      & 2115 & $\pm$ & 7 & 26  & $\pm$& 14 & - & 7.28  & $\pm$ & 0.07 & 8.43  & $\pm$ & 0.46  \\
B904 & e      & 2205 & $\pm$ & 2 & 14  & $\pm$& 4  & - & 7.55  & $\pm$ & 0.06 & 7.96   & $\pm$ & 0.26 \\
\hline
\end{tabular}
\label{table:MLRatios}
\end{center}
Column(2) shows the morphological classification of the dwarf sample as indicated in the FDS catalogue. Column(3)\&(4) are the kinematics derived in this paper.  Column(5) shows the aperture corrected specific angular momentum measured in paper I. In column (6) the stellar mass is measured following \citet{Taylor:2011} as in eq.3. Column (7) shows the measured dynamical masses following \citet{Wolf:2010} in eq.9.
\label{table:kinematics}
\end{table*}

\section{Discussion}
\label{sec:discussion}
In this work, we present the analysis of the kinematic scaling relations and the kinematic and DM properties of a statistically significant sample of 38 dwarf galaxies in the Fornax cluster, the most significant sample of low-mass early-type dwarf galaxies in a cluster to date. As shown in Paper 1, this sample represents the dwarf galaxy population in the Fornax cluster above $M_*\approx10^{7}M_\odot$. By probing such low stellar masses, this sample provides the chance to trace the transition regime between low and high mass galaxies, as our sample is the first dwarf galaxy sample with substantial overlap in stellar mass to the well known sample of Local Group dwarfs (see e.g. top panel of fig.~\ref{fig:ML}).

\subsection{FP and Stellar Mass Fundamental Plane}

Faint low mass dwarfs, with their low stellar density, are sensitive to internal mechanisms such as feedback through star formation and supernovae, and external environmental mechanisms such as ram-pressure, tidal stripping, or galaxy-galaxy harassment. So it is expected that a combination of effects shaped the internal structure of dwarf galaxies. To obtain a better understanding of how these processes affect their structure, their dark matter fraction and stellar populations, we study the deviation and scatter of galaxies not just from the FP, but also from a more sophisticated two-dimensional plane, the Stellar Mass Fundamental Plane ($M_*$P). 

We analyzed two projections of the FP, defined each in two different ways. When defining the FP using the giants of FB11, we find that the DSAMI early-type dwarf galaxies are offset from the plane, with a fainter surface brightness for a given effective radius, consistent with T14. We find that our fainter dwarfs are deviating more from the FP than the objects of T14.

When using all objects to define the FP, we find that the more massive dwarfs lie slightly below the plane in the (log $\sigma$ projection), while the low mass dwarfs lie above it. Correcting for the effects of stellar populations, we find a similar behavior on the Stellar Mass Fundamental Plane. The more massive dwarfs lie slightly (1$\sigma$) below the plane, while the low mass dwarfs lie significantly (3$\sigma$) above the plane. 

\begin{figure}
   \centering
   \includegraphics[width=8.3cm]{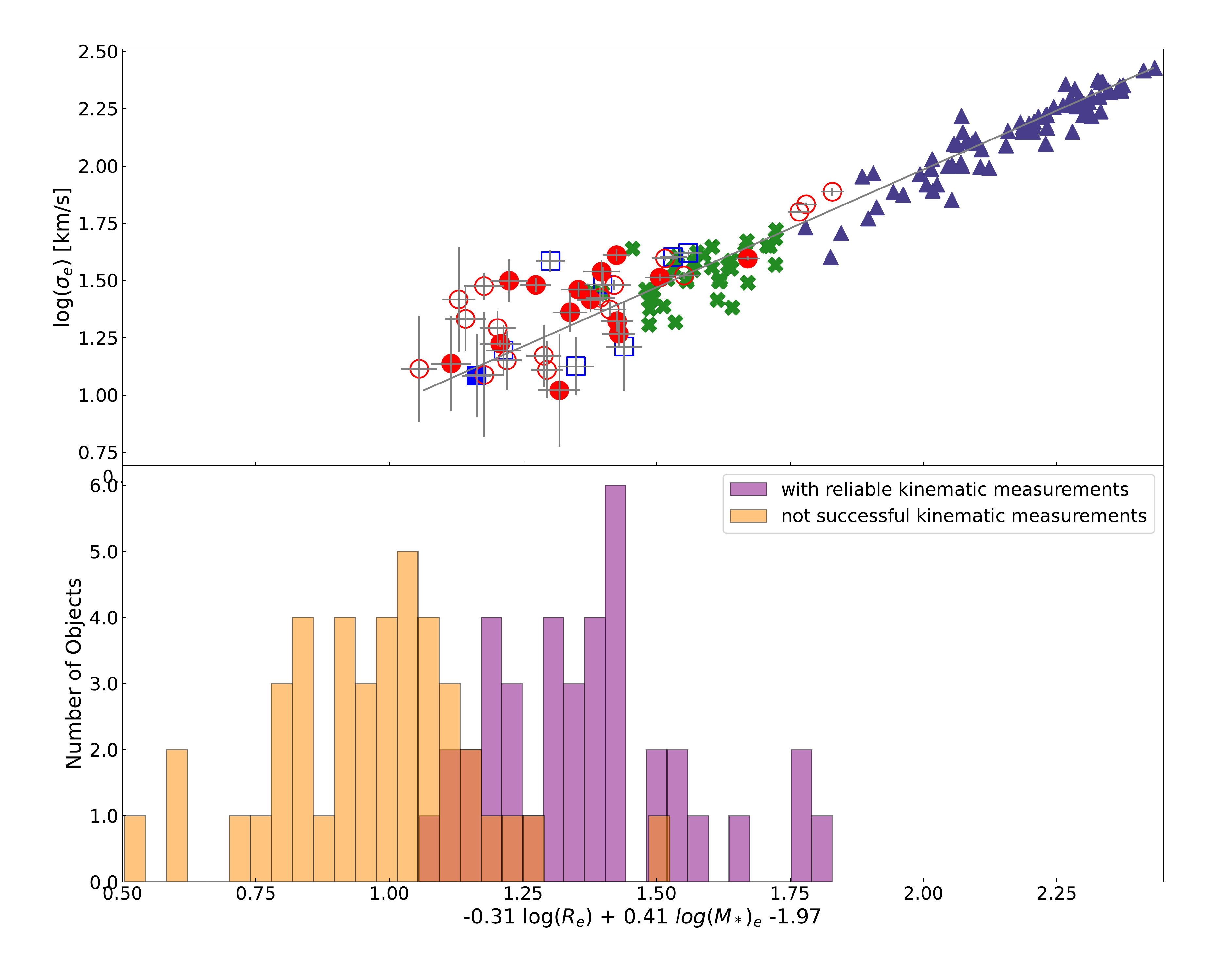}
         \caption{Top: Stellar Mass Fundamental Plane of the whole sample of galaxies in Figure\ref{fig:FP-Mass}. Below:  histogram of the number of dwarf galaxies observed in 2015, 2016 and 2018. Purple bars indicate the number of galaxies with successful kinematic measurement, while orange bars point to the galaxies observed for which the S/N was not large enough or the galaxy was too faint for reliable kinematics to be extracted from its spectra. See also Appendix B. }
         \label{fig:MassPlane_hist}
\end{figure}

It is worth noting that the observed strength of this 'flattening' of the M$_*$P may be affected by the observational limits of our sample. As can be seen in Fig.~\ref{fig:MassPlane_hist}, many of the fainter targets in our original sample did not yield high enough S\/N in their spectra for us to reliably measure their velocity dispersion, or that their velocity dispersions were lower than 10 km/s (see the simulations in Appendix B). 

\citet{Toloba:2012}, by studying the FP, conclude that the location of dEs with respect to the plane is mainly related to their dynamical mass-to-light ratio, which can be used as an estimate of the dark matter content of dEs relative to Es. In contrast, \citet{deRijcke:2005} suggest that different star formation histories can explain this offset of dwarfs from the plane of massive ellipticals. With the stellar-mass plane we can now resolve the issue, showing that independent of stellar population effects faint early-type dwarfs still lie above the M$_*$P, indicating that dynamical mass-to-light ratios are, most likely, changing.

In Figures \ref{fig:FP-FB}, \ref{fig:MP-FB}, \ref{fig:FP-LTS}, and \ref{fig:FP-Mass}, a combination of scatter and box-whisker plots are included. These plots represent the relation between the perpendicular deviation of the galaxies from the corresponding plot and the dynamical-to-stellar-mass ratio of galaxies. The measurements of dynamical masses follow the \citet{Wolf:2010} formula of equation 9. In the scatter plot, objects are grouped into giant early-types ($9.5 < log(M_*/M_{\odot}) \leq 11.5$, purple circles), massive dwarfs ($8.5 < log(M_*/M_{\odot}) \leq 9.5$, orange circles), and low-mass dwarfs ($7.0 < log(M_*/M_{\odot}) \leq 8.5$, green circles). By design, in all four figures, $D_\bot$ of galaxies is correlated with their DM content. The tightness of this relation (R-value) shows whether the relation is dependent on a third parameter or not. For instance, the R-value in the last subplot of Fig.~\ref{fig:FP-FB} is 0.76, which means the regression line cannot predict the scatter of the points perfectly, and a third parameter contributes to the deviation of objects from the plane. This third parameter is the stellar population of galaxies and is removed when we replace surface brightness with stellar mass in Fig.~\ref{fig:MP-FB}, where the R-value in its last subplot is 0.93. From the mathematical perspective, the perpendicular deviation from the stellar-mass plane is another representation of dynamical mass of dispersion supported galaxies in equation 9, which again shows that the deviation of dwarfs from the $M_*$P of giants is a robust estimate for their DM content. Similarly, the R-value in Fig.~\ref{fig:FP-LTS} is larger than in Fig.~\ref{fig:FP-Mass}.

The box-whisker plots are used for displaying variations in the three subgroups of giant early-types, massive dwarfs, and low-mass dwarfs of the scatter plot on the right side. In Tables \ref{table:d_rms_Giants_Planes} and \ref{table:FaintBright-FPandMP}, we only study
the (weighted) mean and (weighted) RMS values. Here in box-whisker plots, we study the distribution of deviation of each subgroup. The boxes represent where the middle 50\% (first quartile, median, and third quartile) of the sample sits in, and each whisker shows the min and max while circles present outliers. In all subplots, giant early-types are centered at the tolerance, and dwarfs have excess variations.

\subsection{Dynamical Mass vs. Stellar Mass}

In the analysis of the ratio of dynamical and stellar mass, we use the samples discussed in this paper, as well as a sample of dwarf galaxies in the Local Group by MC12. In Table~\ref{table:M_Dy/M_St}, we saw that the mass-to-light ratio increases for low-mass dwarfs. This behavior is seen for the SAMI dwarfs, the ones of T14, and the objects in the Local Group. This pattern is also seen in \citet{Penny:2016} and in \citet{Tortora:2016} although our sample here is larger and goes down to much lower masses.
It looks as if the relationship in Fig.~\ref{fig:ML} is not a simple power-law, in contrast to the homology assumption in most of the scaling relations. To investigate the fundamental manifold \citep{Zaritsky:2006} in non-linear space further, a sample with a larger stellar mass range is required.

By ignoring MOND, and assuming that our galaxies are in dynamical equilibrium, the reason for the increased fraction of dark matter per stellar mass in Fig.~\ref{fig:ML} is thought to be a truncation/suppression of star formation at early times, which leaves the galaxy with very few stars for its dark matter content. These increased values have been investigated before \citep{Benson:2000, Marinoni:2002, vandenBosch:2003, Zaritsky:2006, Behroozi:2010} and indicates that the effect of feedback becomes prominent in the lower mass dark matter halos. Energy feedback by supernovae and UV photoionization are considered the two main processes responsible for increasing $M_{dyn}/M_*$ with decreasing $\sigma$ by lowering the star formation efficiency in low-mass systems. Supernovae winds removing material in these galaxies are more prominent in low mass galaxies because of their weak potential well \citep{Martin:1999, Dekel:2003}, while the intergalactic UV radiation is ionizing the gas \citep{Babul:1992}.  

\begin{figure}
   \centering
   \includegraphics[width=9cm]{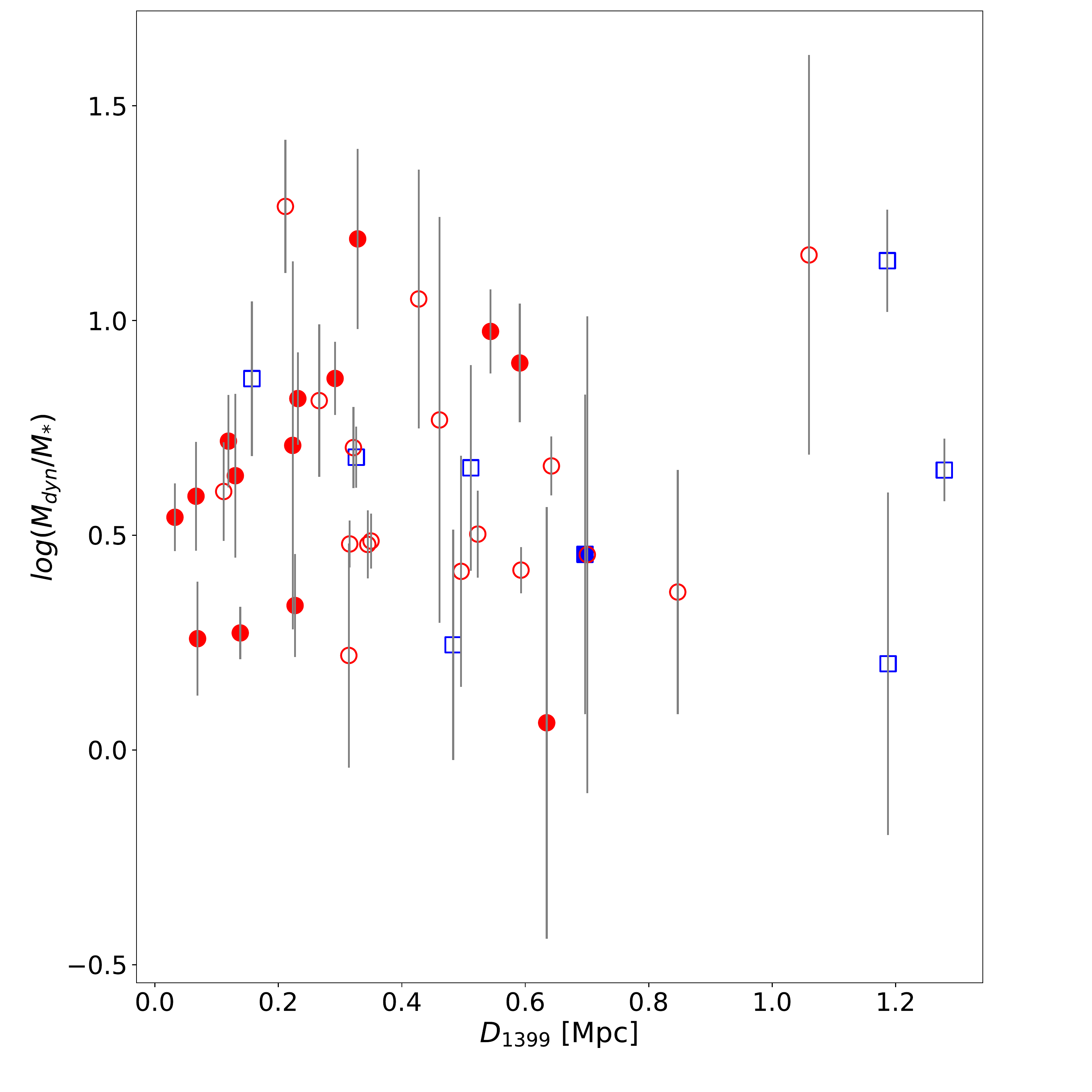}
         \caption{Dynamical-to-stellar mass ratio M$_{dyn}$ / M$_{*}$ vs. projected distance to FCC 213(NGC 1399) in the center of the Fornax cluster for dEs and dIrrs in the SAMI-Fornax sample. The symbols used are the same as in Figure 5.}         
\label{fig:CentricDist}
\end{figure}

When looking at environmental effects, one should note that galaxies of different masses are affected differently by environmental forces. Galaxy harassment results in mass loss, and it has been shown in simulations such as \citet{Moore:1998} and \citet{Smith:2010} that the denser the environment, the higher will be the DM loss in the outskirt of the galaxy compared to stellar mass loss in the central region. Following this claim \citet{Rys:2013} and \citet{Rys:2014}, using a small sample of galaxies in the Virgo cluster, identified a tentative increase of $M_{dyn}/M_*$ as a function of distance from the center of the cluster. It can be seen in Fig.~\ref{fig:CentricDist} that we do not find a similar trend for the radial dependence of the DM content in the Fornax cluster. This agrees with \citet{Penny:2015}, who did not find any trend between $M_{dyn}/M_*$ and distance from M87. The environmental effect can not be induced solely from the current location of dwarfs in the cluster, but also depends on the previous trajectory of the galaxies in the cluster.

Ram pressure stripping is caused by the hot inter-cluster medium as a galaxy enters the cluster and removes its atomic gas, eventually halting star formation. This effect results in the majority of dIrrs existing in the outskirt of the cluster, as can be seen at modest significance in Fig.~\ref{fig:CentricDist}.

\subsection{The tightness of the Stellar Mass Fundamental Plane and galaxy formation}

\begin{figure}
   \centering
   \includegraphics[width=8.5cm]{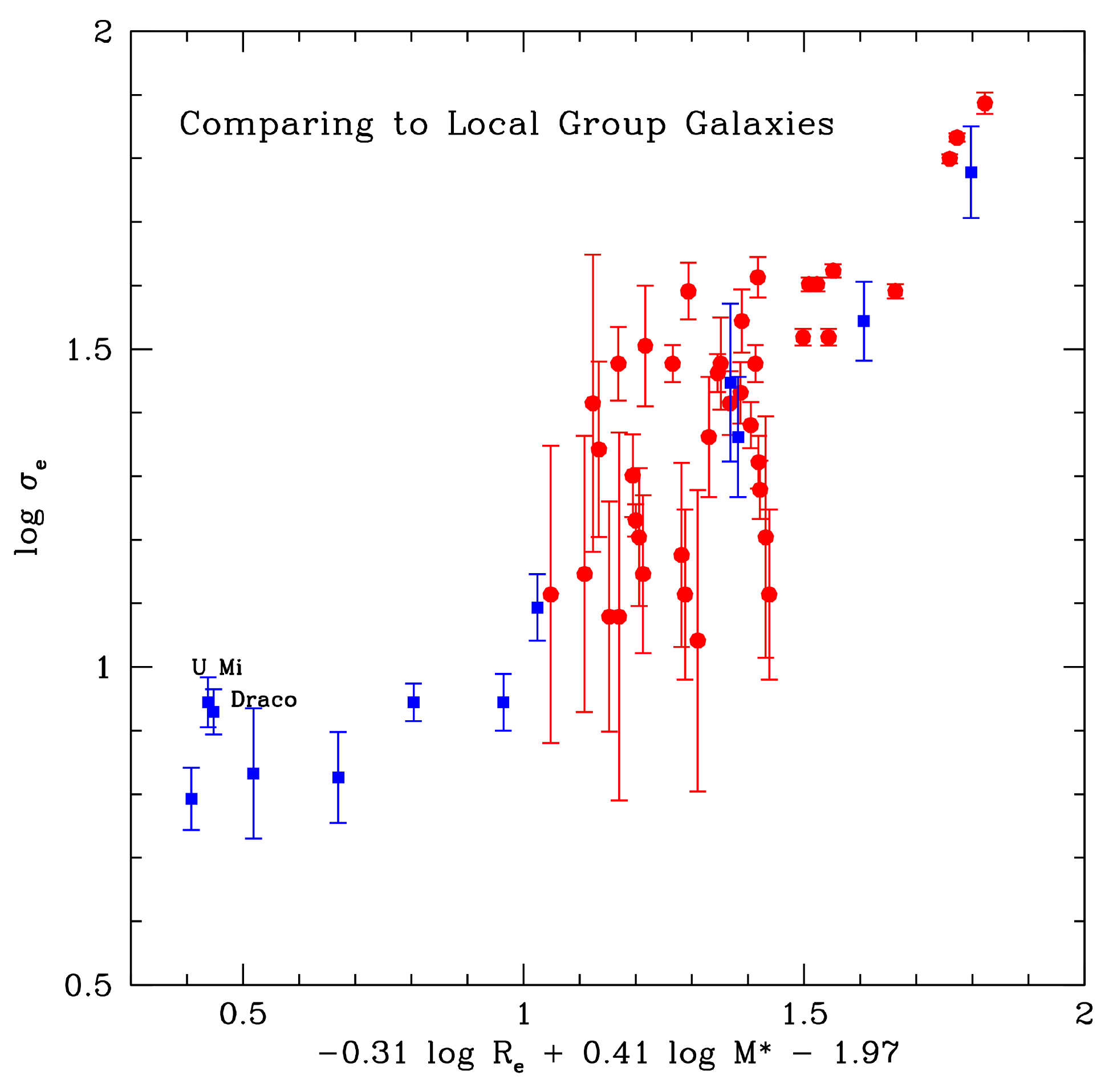}
         \caption{Stellar Mass Fundamental Plane, showing our SAMI-Fornax dwarfs together with 12 Local Group dwarf galaxies, from \citep{Zaritsky:2006}. Here the dwarf sample of T14 and the sample of FB11 are not plotted. The location of both datasets is very similar.}         
\label{fig:FP-LG}
\end{figure}

When plotting the SAMI Fornax galaxies on the Stellar Mass Fundamental Plane together with the early-type dwarf galaxies in the Local Group, using the data compilation of \citep{Zaritsky:2006} (\ref{fig:FP-LG}), we find that both samples overlap well, in the same way that our galaxies overlap with the LG objects in the stellar mass - M/L diagram from the previous subsection.  This seems to imply that the FP does not depend on the environment, it being the same in the Local Group and in a cluster environment, so that the processes that determine the internal structure of galaxies, for example described in the previous subsection, are the same. To determine how this can be the case, and what conclusions can be drawn about the efficiencies of the various processes is not easy, since local densities vary, and environments close to large galaxies like the Milky Way can be similar in some respect to environments in the Fornax Cluster. We therefore leave this discussion to a future paper.

One of the main conclusions of this paper is that galaxies, from dwarfs to giants, fall onto the FP. If one also includes galaxy clusters in this picture (\citet{Zaritsky:2006}) this would mean that there would be one planar relation connecting very massive systems to very small systems, ranging across 10 orders of magnitude in mass. As described in Zaritsky et al. (2006a), if galaxies are homologeous, i.e., which have scaleable structure and no preferred size, and they lie on a planar surface, i.e., the FP, there should be a linear relation between M$_{dyn}$/M* and $\sigma$. This is inconsistent with the predictions of $\Lambda$CDM (e.g., Bullock \& Boylan-Kolchin 2017 and Section 5.2), which show that, in reality, M/L increases for large stellar systems, has a minimum, and, again, increases for small stellar systems as we find in this paper. The way out is probably that the central velocity dispersion does not always measure the total mass of a galaxy, and that the distribution of dark and baryonic matter has to be different. For example, dark matter will probably be more concentrated in the outer parts, as shown by flat rotation curves (Rubin et al. 1980, Bosma 1981). 
About the distribution of the baryonic matter inside the dark matter:  Ferrero et al (2021) discuss how the tilts in various scaling relations change by varying the degree of central concentration of the luminous component of galaxies within their dark matter halo. Such analysis could in principle also be done for dwarf galaxies, although here we are in a very different part of parameter space, where it is also not clear what the galaxy components are, since fainter dwarfs generally do not contain bulges (Su et al. 2021), and have much shallower 'disks' (e.g., Venhola et al. 2019), and are likely much rounder than disks of large spiral galaxies (Venhola et al. 2021).

Another consideration is that the FP relation, and especially the stellar mass plane, is so tight. For homologous galaxies scatter is created by processes that affect the baryonic matter in a different way than  the dark matter, such as winds or ram pressure stripping (\citep{Zaritsky:2006}). They will cause scatter in the M/L, perpendicular to the FP. More detailed work will have to be done to understand what this really means and how the scatter in the FP can be used to constrain galaxy formation models.

\subsection{The RAR of the Fornax dwarf galaxies}

\begin{figure}
   \centering
   \includegraphics[width=9cm]{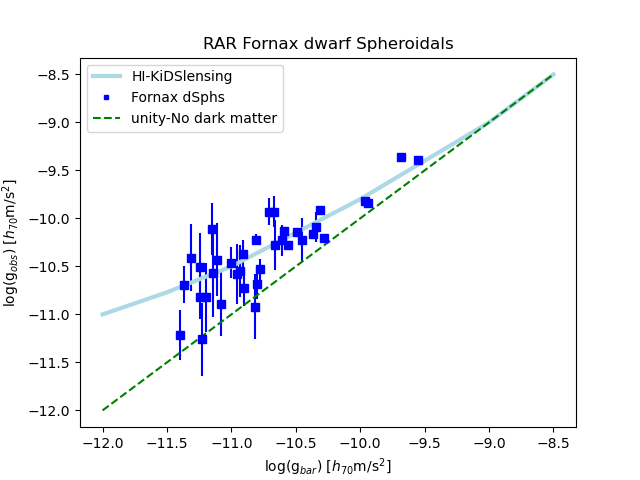}
         \caption{The Radial Acceleration Relation of the Formax dwarf galaxies. The blue line represents the observed relation from HI data and weak lensing. for further details see text.
}         
\label{fig:RAR}
\end{figure}

We next use the empirical Radial Acceleration Relation RAR \citep{McGaugh:2016}, which shows a tight correlation between the acceleration due to the baryonic mass g$_{\rm bar}$,  deduced from optical imaging, and the actual total gravitational acceleration g$_{\rm obs}$, deduced from the extended HI rotation curves of spiral galaxies.  Using a weak lensing analysis this relation has been extended to extremely low accelerations up to hundreds of kpc  distances around isolated  galaxies \citep{Brouwer:2021}.  The  RAR obtained from the lensing  analysis at log(g$_{\rm bar}$) $<$ -11 m~s$^{-2}$  extends closely to the one from the HI data at log(g$_{\rm bar}$) $>$ -11.5 m~s$^{-2}$, both with small error bars, hence settling a narrow empirical relation between the acceleration due to baryonic mass and the total dynamic mass: the sum of the contribution of dark matter and baryons.  Fig.~\ref{fig:RAR} presents a RAR diagram in which the blue line reflects the overall observed and  seemingly universal relation, in this relatively high acceleration regime mostly originating from HI data.

We have computed the data points of the Fornax dwarf sample using  g$_{\rm bar}$ = G M$_*$ / 2 R$_e^2$ and g$_{\rm obs}$ = 3 $\sigma^2 $/ R$_e$ , which reflect the typical accelerations at the effective radii (following \citet{Lelli:2017}).   As the data in Fig 4  demonstrate that the measured  velocity dispersions do not vary much with radius,  the value at R$_e$ seems a fair representation of the overall systems.   Of course this derivation assumes dynamical virialized, stable systems, which is not necessary  always the case. The vertical error bars correspond to measurement errors of the velocity dispersion. For clarity, we did not include the horizontal error bars, which are much smaller. In a future paper uncertainties in the stellar M/L values, for example due to changes in the IMF, will be discussed.
Fig:~\ref{fig:aper-corr} shows that, with the exception of a few galaxies, the Fornax dwarf ellipticals conform the empirical universal RAR.  Hence, their dark matter fraction is following this relation and matches to the outer regions of spiral galaxies with HI detected mostly beyond the optically visible disks, and is higher than the overall DM fraction of Spirals.  A key notion from this graph is that also the observed Fornax dwarf elliptical galaxies seem to follow the tight universal empirical  RAR. This, in fact, confirms  the relation on sub-kpc scales, bracketing the relation on small scales, while it  extends up to 100 kpc scales and beyond. This relation is the empirical basis of MOND, is theoretically predicted by Verlinde’s Entropic Gravity, and reproduced by the MICE LCDM semi-empirical numerical simulations, but not by the LCDM BAHAMAS numerical simulations \citep{Brouwer:2021}. Further advancements will require sensitive observations of fainter dwarf galaxies with at least an order of magnitude lower acceleration regimes. 

\section{Conclusions}
Using multiplexing IFUs with long integrations, we explored the location of a statistically significant sample of extragalactic low-mass dwarf galaxies in the Fornax Cluster on a number of kinematic scaling relations. This study goes down more than a factor 10 in stellar mass than has previously been done, for a considerable sample.
To facilitate the interpretation of the results, we introduced the Stellar Mass Fundamental Plane, in which the effect of different stellar populations is corrected by replacing surface brightness by stellar mass.  We also introduced the perpendicular deviation of dwarf galaxies from the $M_*$P defined by giant early-types as a robust estimate for their DM content.

Our main result is that the lowest mass dwarfs ($10^{7}-10^{8.5}$ M$_\odot$) deviate considerably from both the Stellar Mass FP and the  FP. In addition, measuring the dynamical mass of these faint, dispersion-supported galaxies in a simple way also shows an increase in the mass-to-light ratio for low-mass galaxies when compared to brighter dwarfs and giants.  This observed higher DM content of low-mass galaxies seen in the Stellar Mass Fundamental Plane and in the relation between $M_{dyn}$  and $M_*$ is consistent with the high M$_{dyn}$/L ratios seen in Local Group galaxies. It shows that, most likely, star formation must have been inhibited in the early phases of galaxy formation. 

We find that the position of our galaxies on the Stellar Mass FP agrees with the galaxies in the Local Group, using the data of \citep{Zaritsky:2006}. This seems to imply that the position of dwarf galaxies on the FP depends on the environment in the same way.  We also find that the Stellar Mass FP is a tight relation, indicating that these objects are well virialised.

Additional results are that on the stellar mass plane we see that dIrr are found in the same location as dEs. This possibly means that the internal structure (and also M/L ratios) in star forming dwarfs is very similar to quiescent dwarfs. This might not be surprising, given their similar structural parameters (e.g. Venhola et al. 2019). We do not see any dependence of the M/L ratios on nucleation, nor as a function of distance to the Fornax cluster center.

\section{DATA AVAILABILITY}

The reduced data underlying this article is available on request 1 year after publication of this paper. The raw data is publically available in the AAT data archive.
  
 \section*{Acknowledgements}
We would like to thank Francesco D’Eugenio for handy suggestions in regression analysis.  FSE acknowledges the funding support by the ESO Ph.D. studentship program. RFP acknowledges the financial support from the European Union’s Horizon 2020 research and innovation
program under the Marie Sklodowska-Curie grant agreement No. 721463 to the SUNDIAL ITN network. NS acknowledges the support of an Australian Research Council Discovery Early Career Research Award (project number
DE190100375) funded by the Australian Government and a
University of Sydney Postdoctoral Research Fellowship. JJB acknowledges the support of an Australian Research Council Future Fellowship (project number FT180100231) funded by the Australian Government. GvdV acknowledges funding from the European Research Council (ERC) under the European Union's Horizon 2020 research and innovation programme under grant agreement No 724857 (Consolidator Grant ArcheoDyn). J.~F-B  acknowledges support through the RAVET project by the grant PID2019-107427GB-C32 from the Spanish Ministry of Science, Innovation and Universities (MCIU), and through the IAC pro
ject TRACES which is partially supported through the state budget and the regional budget of the Consejer\'ia de Econom\'ia, Industria, Comercio y Conocimiento of the Canary Islands Autonomous Community.




\bibliographystyle{mnras}
\bibliography{fornax_dEs} 



\newpage
\appendix

\section{Resolution of SAMI}
 
When extracting information from a spectrum, the instrumental resolution is a critical factor in obtaining accurate measurements. To get a precise estimate of the resolution, we use stars observed with the SAMI instrument. We have ten stars, which have negligible velocity dispersions compared to the instrumental resolution. The spectral lines are widened by the instrument's resolution, and we measure the instrument's resolution by calculating the spectra's velocity dispersion.\\

 \begin{figure}
  \centering
  \includegraphics[width=9cm,height=7cm,keepaspectratio]{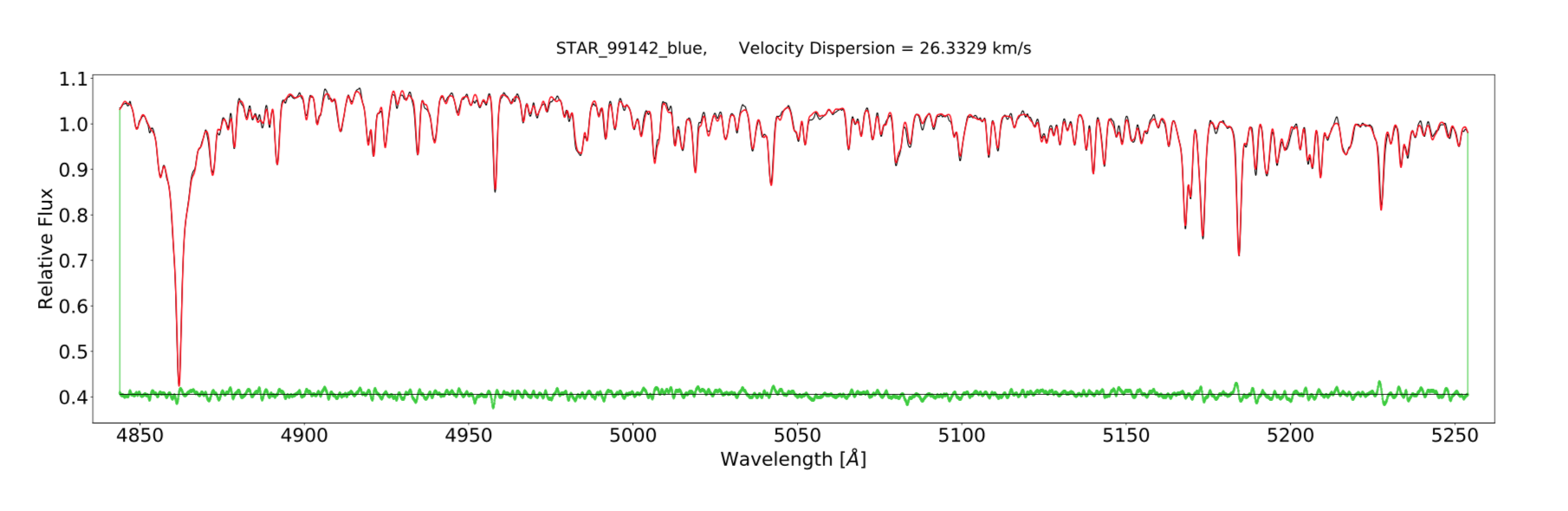}
  \includegraphics[width=9cm,height=7cm,keepaspectratio]{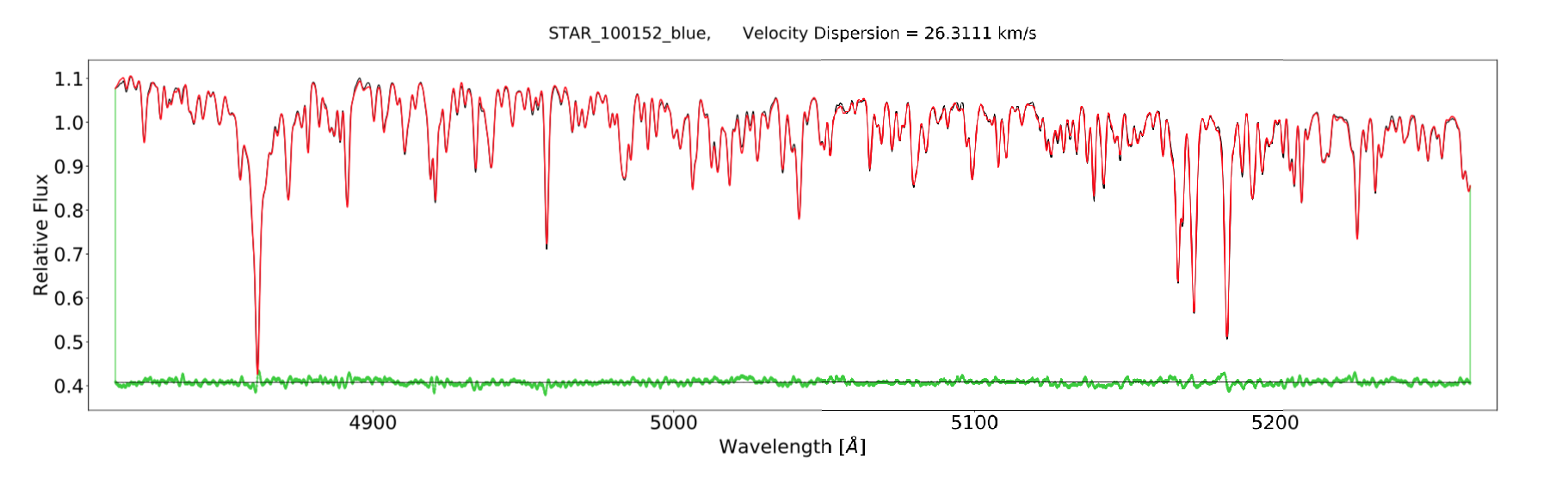}
         \caption{Two examples of the pPXF best-fits of stars observed by SAMI instrument. The black spectrum is the spectrum of the star, red is the best-fit spectrum derived by pPXF, and the residuals are plotted in green.}
         \label{fig:stars_best-fit}
\end{figure}
\begin{figure}
  \centering
  \includegraphics[width=15cm,height=7cm,keepaspectratio]{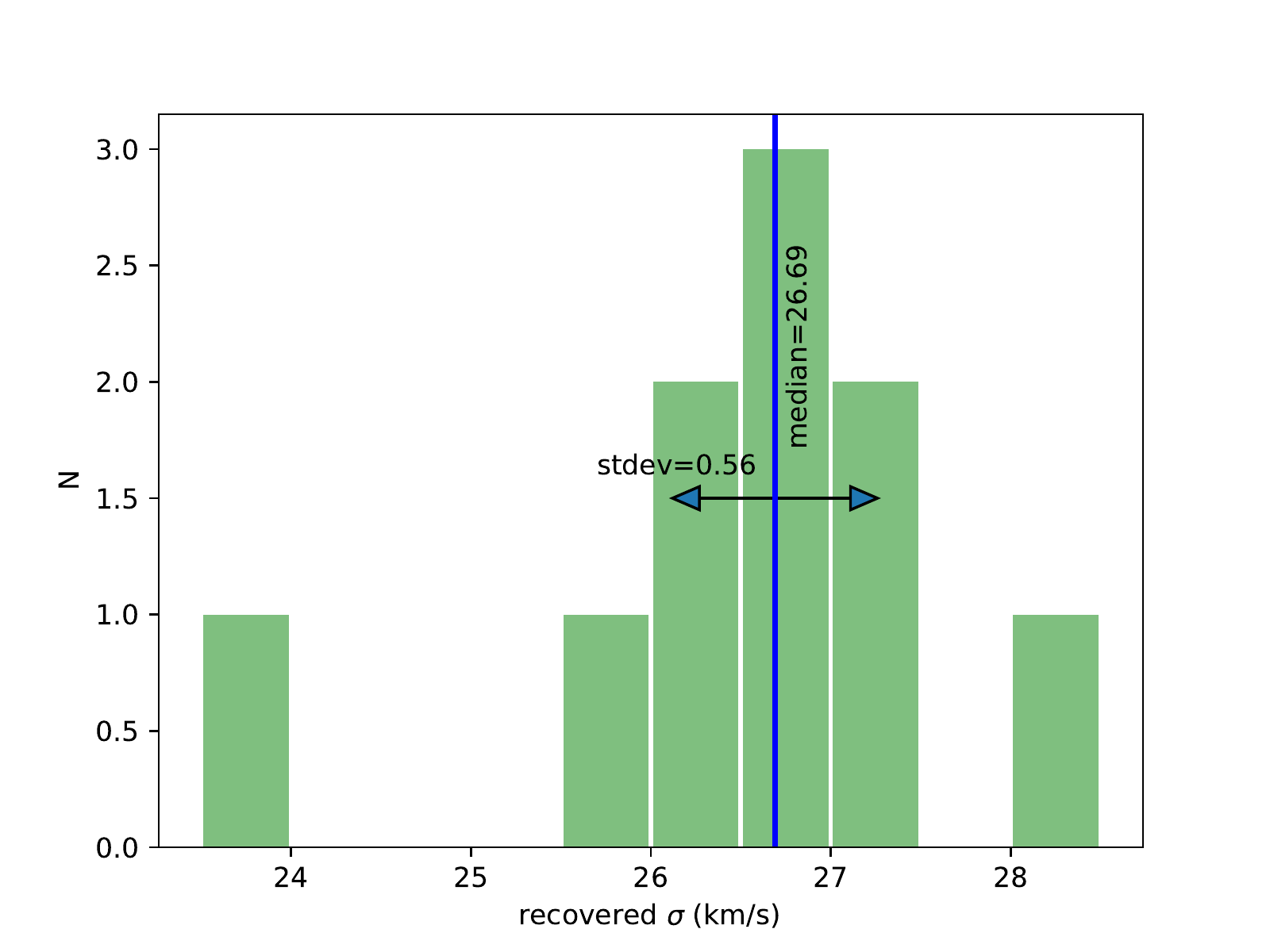}
         \caption{pPXF recovered sigmas of the stars observed with the SAMI instrument. After sigma clipping the resolution of the instrument is 26.69 km/s with standard deviation of 0.56 km/s. }
         \label{fig:sigma_clip}
\end{figure}

The process of determining the spectral resolution goes as follows. We first rebinned the reduced spectra to logarithmic pixel scale (here ~ 8 km/s). To get the best-fitting spectrum, we use the ELODIE \citep{Prugniel:2007} stellar library in its low-resolution release of R=10000  (spectra with FWHM of $0.55\AA $). Another important point is that instead of using the whole stellar sample of ELODIE ($\sim2000$ stars), we are using only about 100 stars with metallicity [Fe/H] ranging from -2.0 to 0.5, gravity log g from 0.5 to 5, and temperature from 2500 to 20000K. The quality of the final best-fits of the stars by pPXF and their corresponding calculated velocity dispersion are somewhat dependent on the star spectrum, the night of observation, and the pointing and the focus of the telescope at that night. We apply additive polynomials of order four and no multiplicative polynomial in the pPXF. In Fig. \ref{fig:stars_best-fit} two of our ten-star spectra and their corresponding best-fits are shown. 

To remove outliers, we use sigma-clipping, in which values smaller or larger than a specified number of standard deviations from the central value are rejected. In our case, two of the values are rejected in the first iteration as they are situated outside the median $\pm$ \noindent two standard deviations of the original data set. There is no need for further iterations as the relative reduction in the value of standard deviation is small. So from the ten calculations of the SAMI instrumental resolution, eight values are used, which gives the following estimate of the resolution
\begin{eqnarray}
\sigma_{SAMI}=26.69\pm0.57(km/s)
\end{eqnarray}
In Fig. \ref{fig:sigma_clip}, the distribution of the calculated $\sigma$'s of the ten stars is shown. 
The above systematic uncertainty is considered in all of the previous measurements in the paper, also in the simulations in the next section.

\section{Reliability of the measurements}
It is essential to determine the minimum reliable velocity dispersion in our measurements and an approximation of uncertainties inferred from the limited resolution. In particular, as we study faint dwarf galaxies inside the Fornax cluster, we are reaching very small velocity dispersions around 11 km/s. Based on previous studies such as \citet{Toloba:2012}, pPXF measures extra broadening down to $0.4\sigma_{resol}$ which means around 10. km/s in our case. To ensure this minimum reliable value, we carry out a simulation using a Monte-Carlo method to determine the recovered velocity dispersion by pPXF.\\ As for the simulated galaxy, we use the library of Single Stellar Population (SSP) models of PEGASE.HR  \citep{LeBorgne:2004} with an FWHM of $0.55$\AA\ and specified ranges of age and metallicity. Then we go to logarithmic space with a velocity scale of 8 km/s and convolve the spectra to the SAMI instrument's resolution as we want a galaxy spectrum observed by SAMI, which means that each line will be broadened to $\sigma_{\rm SAMI}$ $\sim$ 26.69 km/s. The second convolution is the input velocity dispersion (e.g., 15, 20 km/s) of the simulated galaxy. Finally, we add a random noise from a normal distribution with various levels of S/N. Noise is added pixel by pixel dependent on the pixel's corresponding flux and the considered S/N. For each simulated galaxy spectrum, we use pPXF, the same as we did for our SAMI dE galaxy sample. For getting the best-fitted spectrum of each simulated galaxy with pPXF, we need a stellar library for which the same $\sim100$ ELODIE stellar template as the previous section is used. They will be logarithmically rebinned to the same spectral pixel (8 km/s) and convolved to the galaxy's instrumental resolution. To accurately estimate the recovered sigma, each spectrum is tested multiple times ($\sim$100) at a given input sigma and S/N.

\begin{figure}
   \centering
   \includegraphics[width=9cm,keepaspectratio]{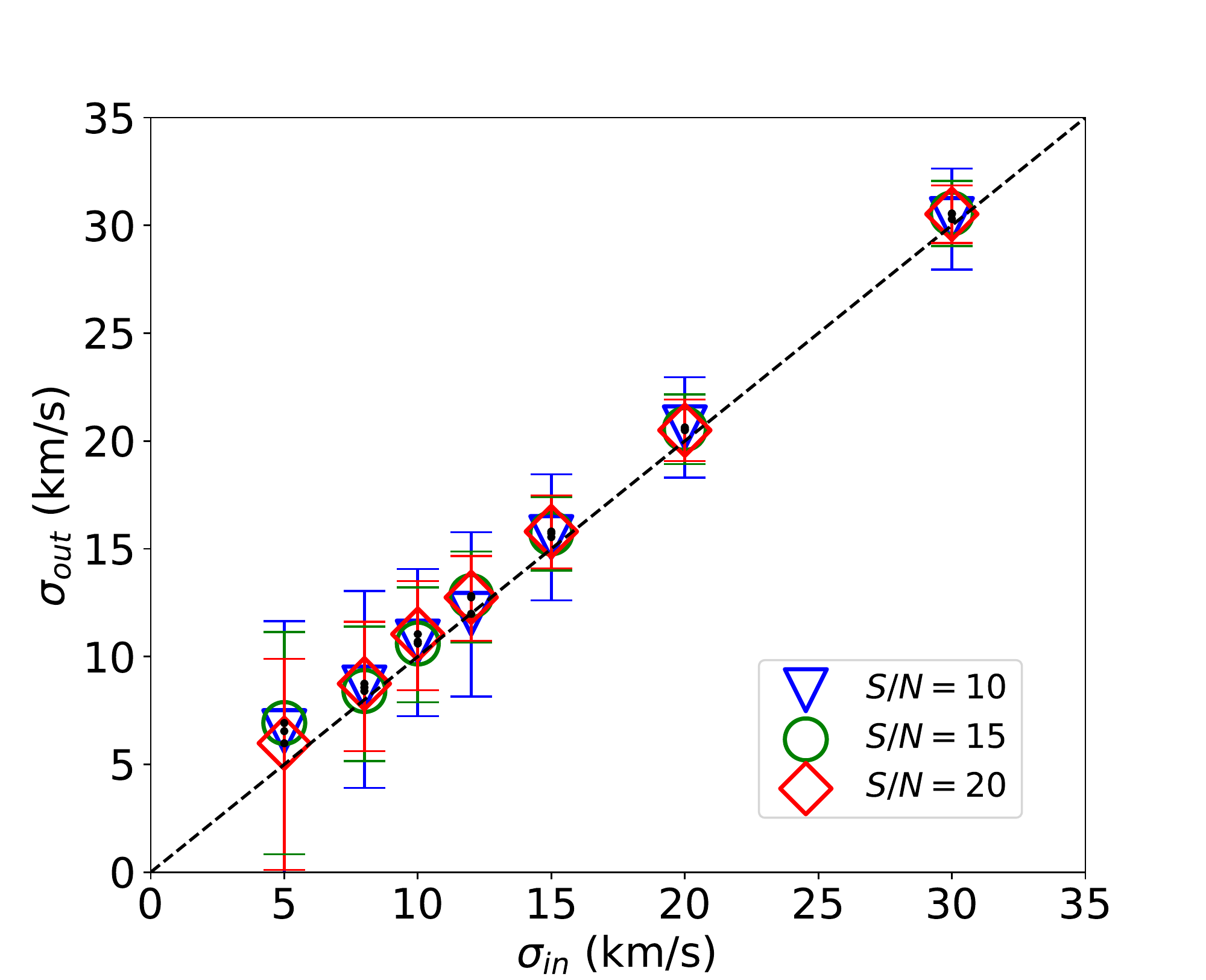}
         \caption{Comparison between the velocity dispersion of a simulated galaxy ($\sigma_{in}$ (km/s)) with t=4Gyr and [Fe/H]=-0.4, and the recovered value by the pPXF ($\sigma_{out}$ (km/s)). Three cases of S/N=10, 15 and 20 are shown with different symbols. The uncertainty of $\sigma_{out}$ is increased in lower sigma spectra while increasing the S/N will be significantly helpful. }
         \label{fig:sigmaIn_sigmaOut}
\end{figure}

In Fig. \ref{fig:sigmaIn_sigmaOut}, we examine the smallest velocity dispersion that can be recovered with acceptable accuracy by using simulated galaxies of t=4Gyr and [Fe/H]=-0.4. As this work is on dEs with small velocity dispersions, we concentrate on small values $\sigma=$ 5, 8, 10, 12, 20, and 30 km/s. Also, at each sigma, the influence of the S/N (=10, 15, or 20) on the certainty of the measurements is investigated. As expected, the uncertainties increase dramatically at the lowest $\sigma$ values (e.g., 5 \& 8 km/s), almost independent of S/N.  The uncertainties at higher velocity dispersions (e.g., 10 \& 12 km/s) depend significantly on the S/N of the galaxy spectrum. For this specified age and metallicity we could recover $\sigma=10$ km/s at S/N=15 with $20-30$\% accuracy and $\sigma=12$ km/s at S/N=15 with $10-20$\% accuracy. For higher velocity dispersions, the accuracy increases considerably, as long as the S/N is larger than 10.

\begin{figure}
   \centering
   \includegraphics[width=10cm,keepaspectratio]{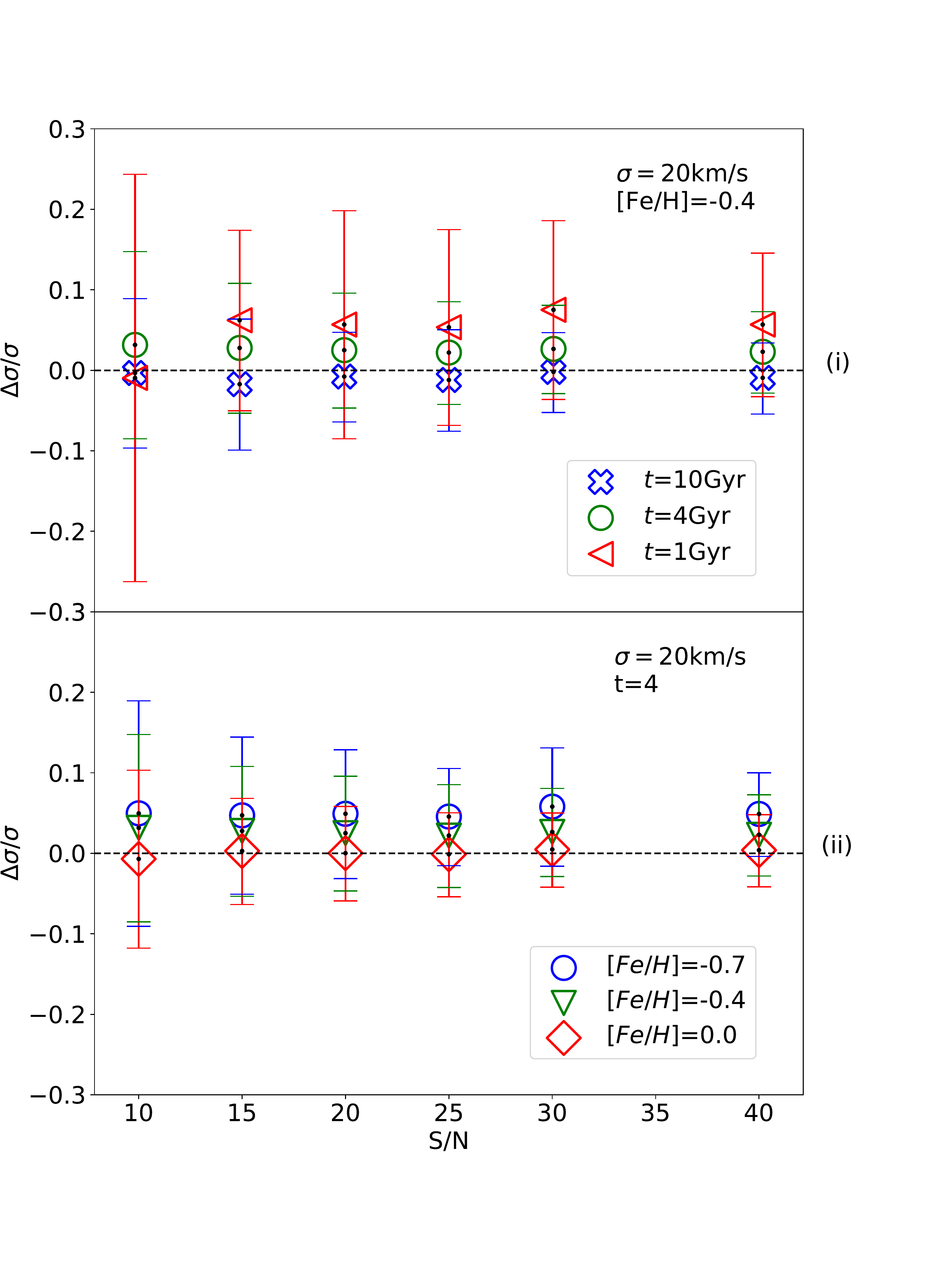}
         \caption{Here we check what the influence of the metallicity and age of a galaxy on the accuracy of its recovered $\sigma$. The upper panel shows the effect of age while the lower panel compares cases with different metallicities at fixed age. The dependency on the age of a galaxy is significantly higher than the influence of the metallicity.}
         \label{fig:fixed_metal_age_sigma}

\end{figure}

\begin{figure*}
   \centering
   \includegraphics[trim={0 10cm 0 1cm}, height=23cm]{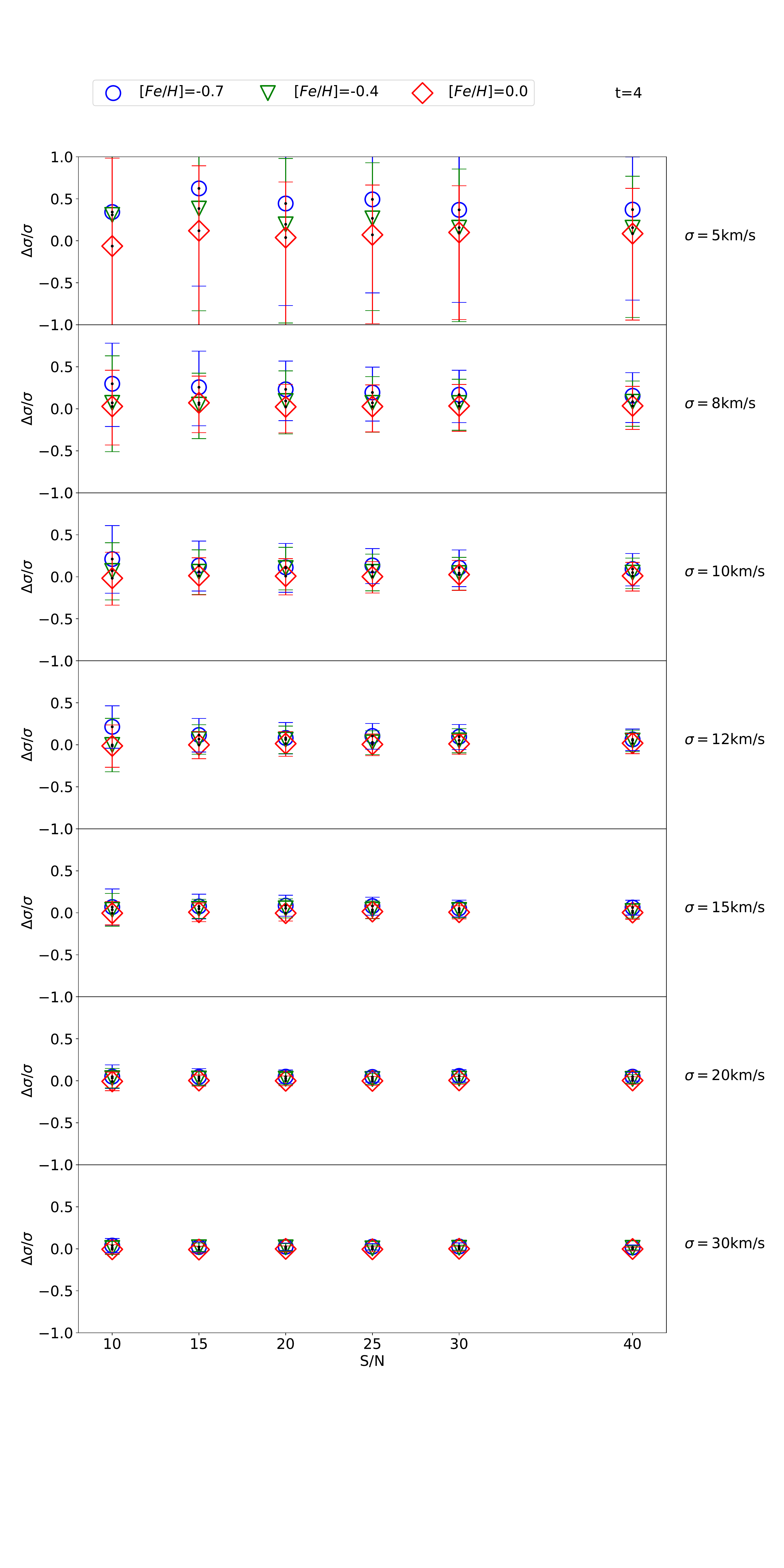}
         \caption{In detail look at the influence of $\sigma$, S/N and metallicity of the galaxy on the accuracy of the recovered $\sigma$ by pPXF.}
         \label{fig:fixed_age}
\end{figure*}

The above procedure was carried out for various input spectra with different ages (t=1, 4, and 10 Gyr) and metallicities ([Fe/H]=-0.7, -0.4, 0.0) to study the influence of the stellar populations on the measurement of the velocity dispersion. The relative differences between the recovered sigmas and the sigmas introduced in the simulated galaxies as a function of S/N, are shown in Fig.~\ref{fig:fixed_metal_age_sigma}(i) for a fixed metallicity ([Fe/H]=-0.4) and in Fig.~\ref{fig:fixed_age}(ii) for a fixed age (t=4Gyr). As mentioned previously, the measurements at S/N=10 result in large uncertainties while, as expected, at higher S/N values, the accuracy improves. In Fig. \ref{fig:fixed_metal_age_sigma}(i) we see a crucial dependency on age especially at age=1 Gyr, for which uncertainties are relatively large. However, our galaxies are probably all older than this value. It might be due to the lack of well-defined absorption lines in young galaxies' spectra. In Fig \ref{fig:fixed_metal_age_sigma}(ii), even though the dependency on metallicity is not as significant as the one on age, there is still an increase in the uncertainties as metallicity gets lower. This is due to metal absorption lines getting weaker.

\begin{figure*}
   \centering
   \includegraphics[trim={0 10cm 0 1cm}, height=23cm]{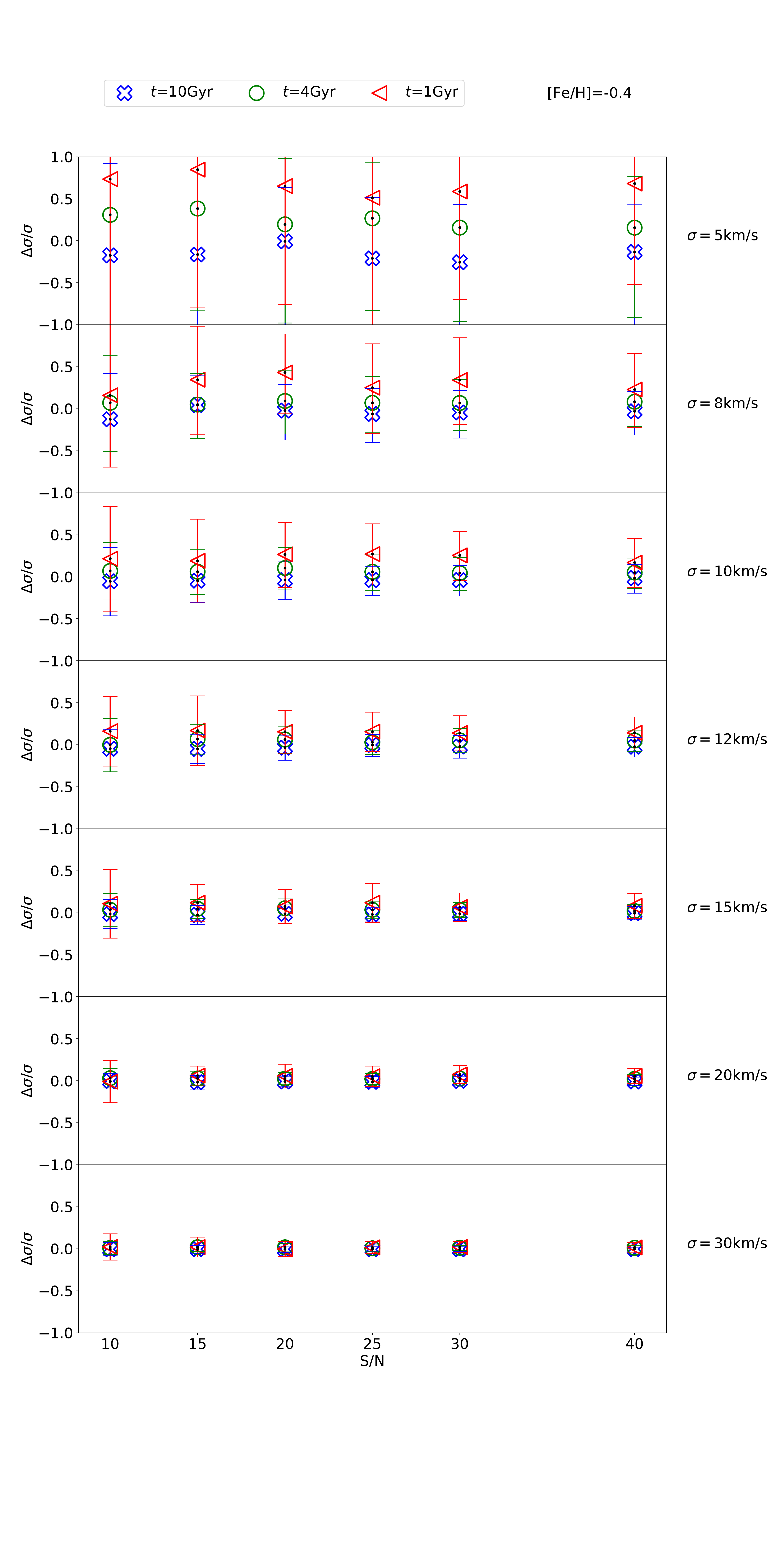}
         \caption{In detail look at the effect of $\sigma$, S/N and age of the galaxy on the accuracy of the recovered $\sigma$ by pPXF.}
         \label{fig:fixed_metal}
\end{figure*}

In Fig. \ref{fig:fixed_age} and Fig. \ref{fig:fixed_metal}, we analyze the influence of the velocity dispersion of the galaxy, the stellar populations, and the S/N of the spectrum in one look. For small velocity dispersions ($\sigma < 10$) large offsets are found even at high S/N (e.g. S/N = 30 or 40). The stellar population does not have an appreciable influence on the measurements, so the velocity dispersions are derived in this $\sigma$ range. On the other hand, we do not find any statistically significant offset for $\sigma>20$ km/s for all S/N,  metallicity, and age values, except for measurements at S/N=10. The accuracy of the recovered sigma becomes important for values around 10-15km/s, exactly where our faintest dwarf galaxies are, and the restriction of pPXF and our instrumental resolution comes in.

\section{The influence of Best-fittings methods on FP}
There are many possible ways to get the best fitted plane in the three dimensional space of $<\mu_e>$, $log(R_e)$, $log(\sigma_e)$. Here we investigate the differences between a few often-used methods.\\
The simplest case of regression analysis is the linear regression, in which we try to find the linear model that describes the trend from a series of points represented in a scatter plot. The relationship between the dependent ($z$) and independent ($x, y$) values can be approximated by a straight line or plane ($z=\alpha x+\beta y+\gamma$). Here we look for the parameters that minimize the difference between the linear model and the observed values $z_i$. There are various known fitting methods available, including the following ones.

\subsection{Weighted Least Square Estimates (WLS)}

This least-square problem has been discussed in a number of papers such as \citet{Isobe:1986}, \citet{Feigelson:1992}, \citet{Akritas:1996}, and \citet{Tremaine:2002}.
If we assume that each individual measured value is drawn from a Gaussian distribution with mean $z_0(x_i,y_i)$ and standard deviation $\sigma_i$, the probability of obtaining the observed set of measurements will be 
$P(\alpha,\beta,\gamma) = \Pi(\frac{1}{\sigma_i\sqrt{2\pi}})exp(-\frac{1}{2}\sum[\frac{z_i-z(x_i,y_i)}{\sigma_i}]^2)$,
and maximizing the probability $P(\alpha,\beta,\gamma)$ is equivalent to minimizing  the sum of the square of the differences between in the fitted values and the estimated values. In reality, most of the time, our data exhibits heteroskedasticity, so instead of minimizing the residual sum of squares, we minimize the weighted sum of squares. In our case, we define weights as inversely proportional to the variance of measurement uncertainties.
\begin{eqnarray}
\chi^2=\sum w_i (z_i-\alpha x_i-\beta y_i-\gamma)^2; w_i = \frac{1}{\alpha^2\sigma^2_x+\beta^2\sigma^2_y+\sigma^2_z}
\end{eqnarray}
Even though the Gaussian assumption may not always be precisely valid, the distribution may be considered Gaussian for a sufficiently large number of counts. This method is well suited to extracting maximum information from small data sets. However, the technique should only be used when the estimates of weights are precise; otherwise, an inappropriate weight can skew the results dramatically. This weight equation is valid only when the parameters are uncorrelated. Otherwise, the terms for covariances have to be included in the denominator.

\subsection{Weighted Least Absolute Estimates (WLA)}

WLA is minimizing the weighted sum of the absolute differences between the observed values and the estimated values
\begin{eqnarray}
\chi^2=\sum w_i \mid{z_i-\alpha x_i-\beta y_i-\gamma}\mid; w_i = \frac{1}{\sqrt{\alpha^2\sigma^2_x+\beta^2\sigma^2_y+\sigma^2_z}}
\end{eqnarray}
This method has also been used for deriving the FP, in works such as \citet{Jorgensen:1996}, \citet{Pahre:1998}, \citet{Falcon-Barroso:2011}, and \citet{Cappellari:2013}, as it is relatively insensitive to a few outliers and more robust by treating all parameters symmetrically \citep{Jorgensen:1996, LaBarbera:2010}. However, the statistical efficiency is decreased, and the fitted parameters' uncertainties get a bit larger. Moreover, minimizing the absolute deviation maximizes a Laplacian probability distribution, so WLA assumes a Laplace distribution for the convolution between the intrinsic and measured values. The assumption of the Laplacian distribution is stronger than the Gaussian one.

\subsection{Robust Linear Regression of Cappellari (lts\_planefit)}

\citet{Cappellari:2013} compares the extended least-square method of \citet{Tremaine:2002} in which intrinsic scatter in the relation is allowed, with the Bayesian method of \citet{Kelly:2007}.  They say the differences between the methods in both 2D and 3D space are insignificant, and the two technically different approaches are conceptually equivalent. They also mention adding clipping to the least-square method in which outliers further than from the best-fitted plane are omitted. Furthermore, they conclude that clipping does not always converge, especially in the presence of significant outliers. In the end, they consider the Fast Least Trimmed Squares (Fast-LTS) regression approach of \citet{Rousseeuw:2006} successful and combine it with a best-fitting method that allows for intrinsic scatter.\\
In the LTS method, the aim is to find a subset of the data which has the smallest $\chi^2_h = \sum_{j=1}^{h} r_j^2$; where $r_j^2$s are the ordered square residuals for a subset of $N/2<h<N$. In summary, lts-linefit or lts-planefit of Cappellari et al. (2013) are iterative steps in which for a set of points of the smallest $\chi_h^2$ standard deviation is measured, the data set is extended with points up to $2.5\sigma$ from the fitted plane, and the linear fit is repeated. 
\subsection{Bayesian Approach (3D Gaussian Likelihood)}
\citet{Magoulas:2012} introduces a 3D Gaussian model in place of a 2D plan-surface with Gaussian scatter in one direction. They investigate how this 3D likelihood lowers the biases when fitting the FP by accounting for the selection effect (hard and soft limits in variables and censoring) and considering correlated errors. The Gaussian probability distribution generalized to three-dimension will be 
\begin{equation}
  P(X_n)=\frac{exp[-\frac{1}{2}X_N^T(\Sigma+E_n)^{-1}X_n]}{(2\pi)^{\frac{3}{2}}\mid\Sigma+E_n\mid^{\frac{1}{2}}f_n}  
\end{equation}
And so $\chi^2$ for a particular galaxy n can be measured as below
\begin{equation}
    \chi^2=X_n^T(\Sigma+E_n)^{-1}X_n
\end{equation}
This measures a galaxy's departure in the best-fitted plane from the 3D Gaussian model.

We have applied all of the mentioned methods on the three data sets of DSAMI \citep{Scott:2020}, T14 \citep{Toloba:2014} and FB11 \citep{Falcon-Barroso:2011}. The recovered parameters can be seen in Table~\ref{table:FP-DiffFits}. We see that the four methods give very similar results, but there are slight differences, so it remains important to check what method various papers in the literature use when comparing them.

\begin{table}
\centering
\caption{Different methods of regression in 3D for FP $log(\sigma) = \alpha<\mu_e>+\beta log(R_e)+\gamma$, which they gave us consistent best-fitted coefficients} 
\begin{tabular}{lllll} %
\hline\hline\\ 
\textbf{FP} & $\alpha$ & $\beta$ & $\gamma$ & RMS\\ [0.5ex] 
\hline\\
WLS & -0.19$\pm$0.008& 0.60$\pm$0.0402& 6.50$\pm$0.175& 0.136\\
WLA & -0.23$\pm$0.008& 0.64$\pm$0.0403& 6.42$\pm$0.176& 0.137\\
LTS-planefit & -0.22$\pm$0.007& 0.70$\pm$0.0330& 6.28$\pm$0.150& 0.130\\
3D Gauss & -0.21$\pm$0.005& 0.80$\pm$0.0300& 5.91$\pm$0.140& 0.127\\
\hline
\end{tabular}
\label{table:FP-DiffFits} 
\end{table}

Among them, LTS and Gauss3D show the lowest RMS. Gauss3D requires priors on the angles (or scatters), is agnostic in terms of the plane's orientation, and treats x,y,z equally. The chosen method throughout the work is LTS\-planefit of \citet{Cappellari:2013} for both FP and M*P.

\section{Faber-Jackson Relation}
\label{FJ}

\begin{figure}
\begin{center}
\includegraphics[width=1.0\linewidth]{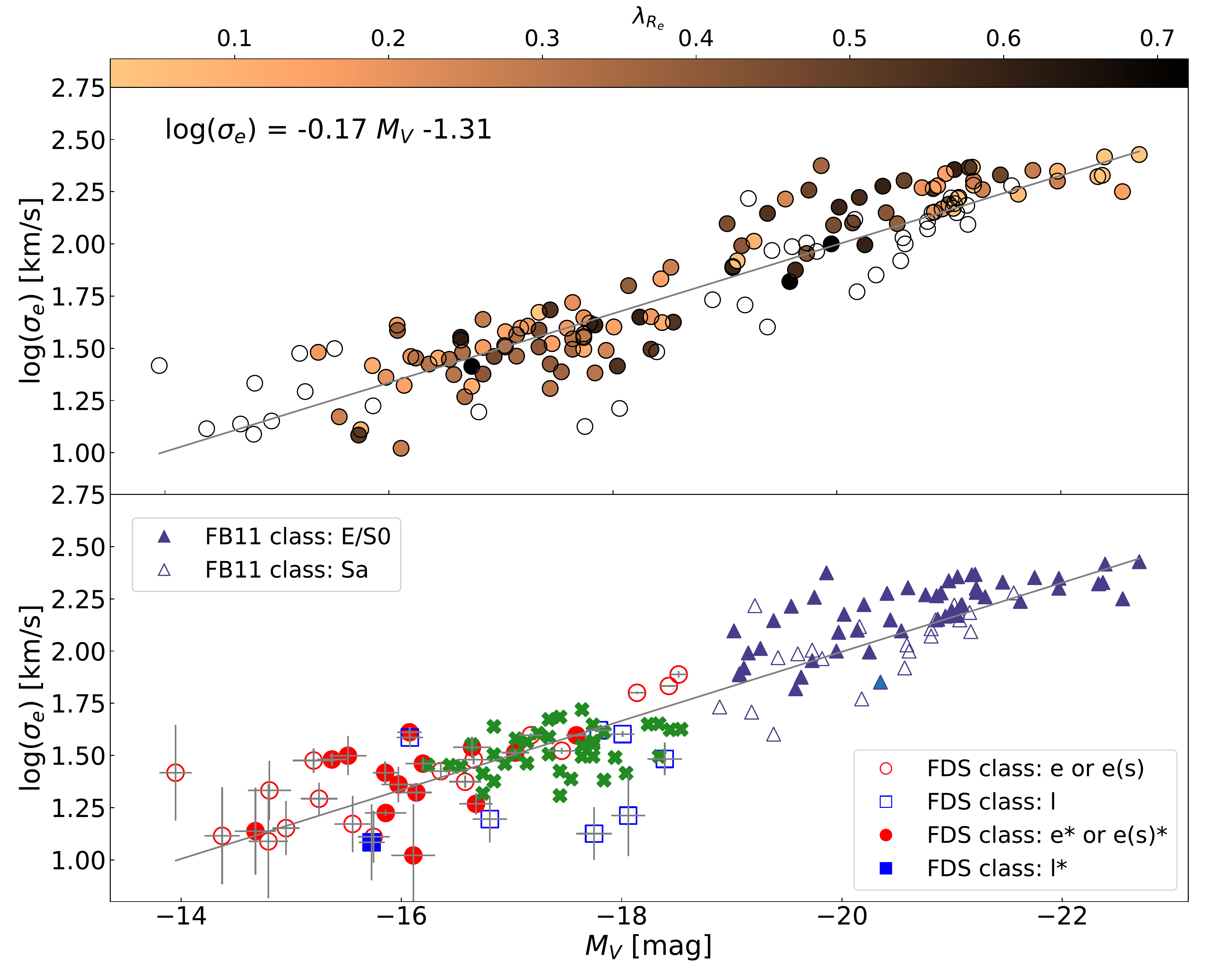}
\includegraphics[width=1.0\linewidth]{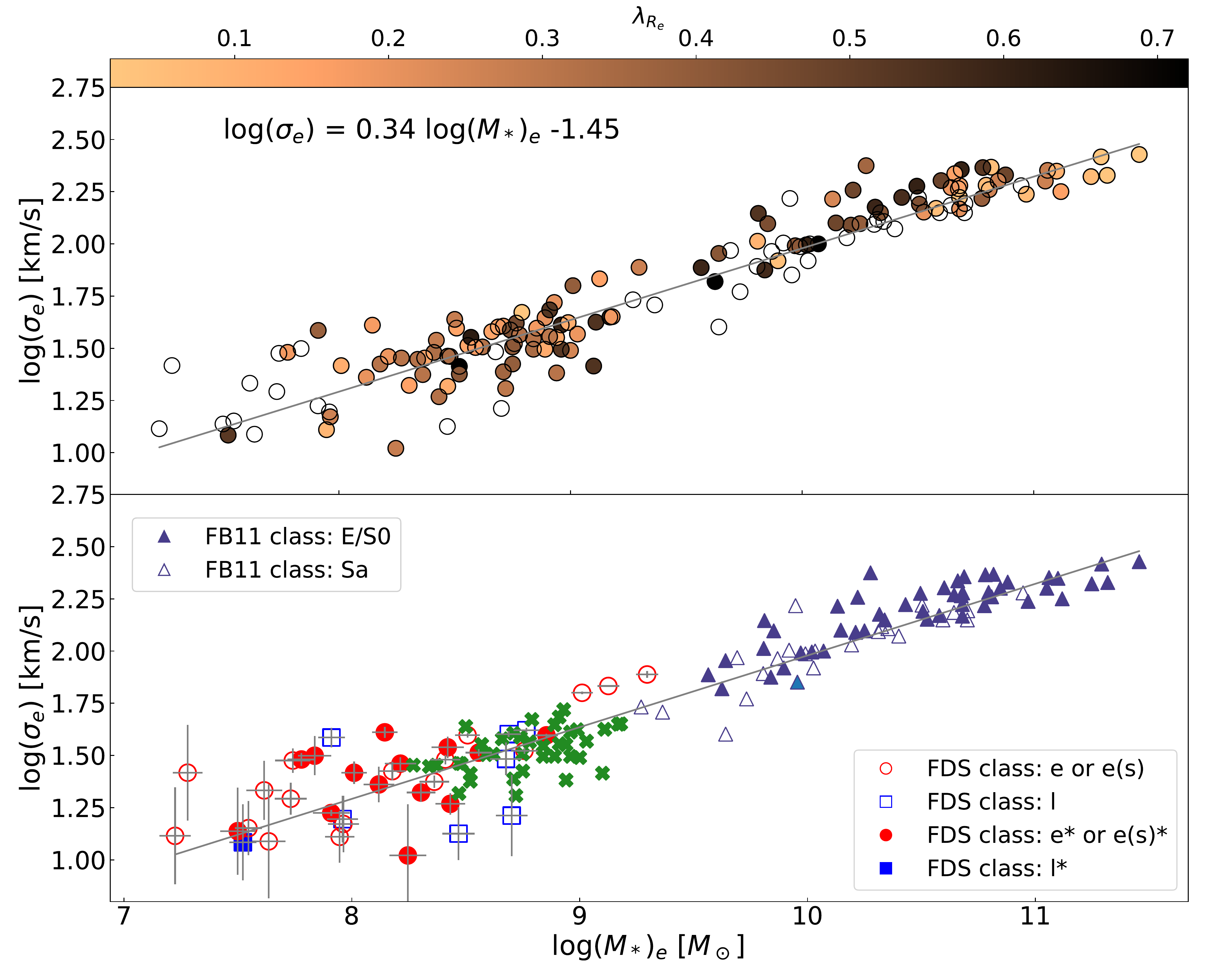}
\end{center}
\caption{	{(a) Faber-Jackson Relation showing the three samples of DSAMI, T14 and F11. In the top panel all of the galaxies are color coded by their specific angular momentum ${\lambda_R}_e$  \citep{Emsellem:2007}. Open symbols indicate galaxies for which no measurements of ${\lambda_R}_e$ are available \citep{Scott:2020}. The symbols in the lower panel are the same as in Fig.~\ref{fig:params(R-M)} (b) Stellar mass version of Faber Jackson relation. The best-fitted relation has been found through weighted least square linear regression. }}
\label{fig:FJ}
\end{figure}

For completeness we also add the Faber-Jackson Relation for this sample. The first discovered kinematic scaling relation for early-type galaxies was the relation between the central stellar velocity dispersion and the luminosity of galaxies \citep{Faber:1976}. This two-dimensional relation $L\propto\sigma^\alpha$, the Faber-Jackson relation, was later shown to be a projection  of the FP. It has been claimed that its slope is dependent on the luminosity range that the relation is fitted to. \citet{Davies:1983} finds $\alpha\sim2$ for low-luminosity elliptical galaxies, and \citet{Schechter:1980} reports $\alpha\sim5$ for luminous elliptical galaxies. By assuming that the mass-to-light ratio and surface brightness are constant, the relation $L\propto\sigma^4$ can be derived from the virial theorem and the gravitational potential of a mass distribution. However, in practice, $\alpha$ will be different, since one does not observe the bolometric luminosity, and galaxies are probably not completely in virial equilibrium \citep{Gudehus:1975}. A comparable method for spiral galaxies is the Tully-Fisher relation, the relation between luminosity and rotational velocity. Both relations are useful in estimating the relative distances of galaxies. 

In Fig.~\ref{fig:FJ}, we show the Faber-Jackson (FJ) relation going down to galaxies of $M\sim 10^{7.2}M_\odot$, about two magnitudes fainter than those of T14. We find that the relation remains tight, towards the faintest galaxies. In the top plot, the galaxies are color-coded by their specific angular momentum $\lambda_R$ \citep{Emsellem:2007} within the effective radius, an indication of being rotationally or pressure supported. The $\lambda_R$ measurements of SAURON galaxies are taken from \citet{Emsellem:2007}, in which Sa galaxies are not included in the fitting, and so are shown by empty circles in the plot. The $\lambda_R$ values of the DSAMI dwarf galaxies have been taken from paper I. Again, the galaxies without $\lambda_R$ measurements are shown with empty circles. These are objects for which only a central $\sigma$-value could be obtained. We find that rotationally and pressure-supported galaxies are indistinguishable on the FJ  relation, and also no dependence on nucleation of SAMI\_Fornax dwarf galaxies is seen.\\ 
The early-type galaxies in our sample are consistent with the sample of T14, that are indicated by green crosses in the lower panel. The galaxies that deviate most from the FJ relation are the dwarf irregulars and some giant spiral galaxies, which is expected, since the FJ relation generally is not meant for studying star forming galaxies. In the lower panel, giant early-type E/S0 galaxies are shown with filled purple triangles and spiral Sa galaxies with empty triangles. The symbols used for dwarfs are the same as in Fig.~\ref{fig:params(R-M)}. Star-forming dwarfs are found below the line (or to the right of it), as they are brighter and younger than quiescent dwarfs. The dependence on stellar populations can be removed by replacing magnitudes with stellar mass as in Fig.~\ref{fig:FJ}, in which the scatter decreases and irregular dwarfs move closer to the line. When using stellar mass instead of luminosity, we see that the scaling relation for dwarfs is the same as for giants. Also, in the overlapping mass range (our sample extends to about 1 decade lower mass), the early-type galaxies in our samples are consistent with the sample of T14. However, the scatter is still significant, which indicates that a third parameter is involved in this universal scaling relation. 

\bsp	
\label{lastpage}
\end{document}